\documentstyle[sprocl]{article}

\input psfig.sty
\def\be{\begin{equation}}
\def\ee{\end{equation}}
\def\bea{\begin{eqnarray}}
\def\eea{\end{eqnarray}}

\def\pslt{p\llap/_T}
\def\eslt{E\llap/_T}
\def\to{\rightarrow}
\def\Re{{\cal R \mskip-4mu \lower.1ex \hbox{\it e}}\,}
\def\Im{{\cal I \mskip-5mu \lower.1ex \hbox{\it m}}\,}

\def\te{\tilde e}
\def\tl{\tilde l}
\def\tf{\tilde f}
\def\tc{\tilde c}
\def\tu{\tilde u}
\def\td{\tilde d}

\def\tb{\tilde b}
\def\tg{\tilde g}

\def\tnu{\tilde\nu}
\def\tmu{\tilde\mu}
\def\tell{\tilde\ell}
\def\tq{\tilde q}
\def\tt{\tilde t}
\def\tw{\widetilde W}
\def\tz{\widetilde Z}

\def\alt{\stackrel{<}{\sim}}
\def\agt{\stackrel{>}{\sim}}
\def\mhf{m_{\frac{1}{2}}}

\begin{document}

\hfill{UH-511-833-95}
\vspace{5mm}
\title{SUPERSYMMETRY: WHERE IT IS AND HOW TO FIND IT
{}~\footnote{Lectures presented at the 1995 Theoretical Advanced
Study Institute, University of Colorado, Boulder, USA}}

\author{XERXES TATA }

\address{Department of Physics and Astronomy, University of Hawaii, \\
Honolulu, HI 96822, USA}


\maketitle\abstracts{We present a pedagogical, but by no means complete,
review of weak scale supersymmetry phenomenology. After a general
introduction to the new particles that must be present in any supersymmetric
framework, we describe how to write down
their interactions with one another as well as with the particles of the
Standard Model. We then elucidate the assumptions underlying the Minimal
Supersymmetric Model as well as the more restrictive minimal supergravity
GUT model with the radiative breaking of electroweak symmetry.
These models serve to guide our thinking about the implications of
supersymmetry for experiments. To facilitate our study of signatures of
supersymmetric particles at high energy colliders, we
describe the decay patterns of sparticles as well as their production
mechanisms in $e^+e^-$
and hadron-hadron collisions. We then discuss how sparticles may be searched
for in on-going
experiments at the Tevatron and at LEP. We review phenomenological
constraints on supersymmetric particle masses from non-observation of any
signals in these experiments, and also
briefly discuss constraints from low energy
experiments and from cosmology. Next, we study new strategies by which
supersymmetric particles may be searched for at supercolliders, and also
what we can learn about their
properties (masses, spins, couplings) in these experiments.
A determination of sparticle properties, we will see, may provide us with clues
about the nature of physics at the ultra-high scale.
After a brief discussion of possible extensions of the minimal framework
and the implications for phenomenology, we conclude with our outlook for the
future.}

\newpage

\section{Introduction and Prelude}\label{sec:Intro}

As Bagger has already discussed in his lectures~\cite{Bagger},
supersymmetry~\cite{Intro,Rev,DPF}
differs from familiar symmetries
such as rotation or Lorentz invariance, gauge
invariance, or the old-fashioned isospin invariance of strong interactions
in one important aspect. Unlike these ``bosonic'' symmetries which relate
properties of a boson (fermion) with those of other bosons (fermions)
a supersymmetry (SUSY) inter-relates the properties of bosons and fermions,
and so, provides a new level of synthesis. The fermionic generator $Q$ of
supersymmetry transforms as a spinor under the Lorentz group. It commutes
with the translation generator. As a result, all states (other than the
zero energy ground state) come in degenerate pairs---for each bosonic state
there is a fermionic state with the same (non-zero) energy. Particularizing
to single particle states, we see that all particles must have supersymmetric
partners (sparticles) with the same mass but spin differing by $\frac{1}{2}$.
Furthermore, because the generator of supersymmetry commutes with internal
symmetry generators, sparticles must have the same gauge quantum numbers as
their ordinary particle partners. Thus, aside from mixing effects, {\it the
gauge
interactions of sparticles are completely fixed.} This is the principal
reason why supersymmetric models have predictive power.

Of course, this also means that unbroken supersymmetry is excluded by
experiment. We know that there are no bosons with, for instance,
the mass and charge of the electron. Such bosons, if they existed, would have
been produced via electromagnetic interactions in accelerator experiments,
and would have been long since
discovered. Supersymmetry must, therefore, be a broken
symmetry. In this case, it is reasonable to ask why we should bother with it
at all. In this context, we must recall an important lesson we have learnt from
electroweak theory: although mass relationships implied by the symmetry may
be badly violated when this symmetry is spontaneously broken, the
underlying relationships between couplings are
nonetheless preserved, with important implications~\footnote{This
does not mean that there are physics implications only
when coupling constant relationships are maintained: knowledge of
how the symmetry breaking terms transform, yields
observable consequences from the underlying (broken) symmetry. The
isospin symmetry of strong interactions (broken by electromagnetic
interactions) and flavour
$SU(3)$ are, perhaps, the best known particle physics examples of this.} for
physics.

The physics of supersymmetry breaking is, however, not yet understood.
Nevertheless, if we believe that the main motivation~\cite{Bagger} for {\it
weak scale} SUSY comes
from the observation that it can protect scalar masses from large radiative
corrections, and thus ameliorate the fine-tuning problem of the SM, we must
also accept that SUSY breaking interactions must be ``soft''; {\it i.e.} they
do not
reintroduce the quadratic divergences that SUSY was introduced to eliminate in
the first place.
Fortunately, the dimensionless couplings of sparticles to gauge bosons and
their
supersymmetric counterparts are not soft, so that the predictive power of SUSY
models
referred to above is unaffected by the introduction of soft SUSY-breaking
terms.~\footnote{That this must be so can be
simply inferred if we recall that the internal quantum numbers completely fix
the
gauge interactions of particles in
any field theory, regardless of supersymmetry.}
Differences between various low-energy models arise in the choice of the
superpotential
function~\cite{Bagger} which gives rise to Yukawa interactions, as well as
assumptions about
the soft SUSY-breaking terms.

The plan of these lectures is as follows. We discuss generalities about
supersymmetric particles
and their interactions in Sec.~\ref{sec:framework}. The assumptions underlying
the Minimal Supersymmetric Model and
the popular supergravity framework extensively used for phenomenology are
discussed in the following two
sections. Supersymmetric model-building~\cite{DREESDPF} is beyond the scope of
these lectures.
We will only focus on those elements that are necesary for our understanding of
the phenomenology.
In Sec.~\ref{sec:decays} and Sec.~\ref{sec:prod} we describe sparticle decay
patterns and production mechanisms,
respectively. Sec.~\ref{sec:sim} briefly outlines the available computer
programs for the simulation of supersymmetric
events at colliders. Empirical constraints on sparticle masses are discussed in
Sec.~\ref{sec:CONSTRAINTS}.
In Sec.~\ref{sec:Future} we investigate how supersymmetry might be discovered
at future colliders~\footnote{A comprehensive
numerical discussion of the branching fractions, production cross sections and
signal cross sections would
require too many figures and tables and it is not possible to
include all these here. We will, therefore, not make an attempt to display
numerical results in these notes, but provide
the reader with an extensive bibliography to the literature where these may be
found.} while the following
section focusses on how we might determine the properties (masses, spins,
couplings) of sparticles. In Sec.~\ref{sec:nonmin}
we briefly discuss possible extensions of the minimal framework, and their
impact on SUSY phenomenology.
Finally, we conclude with a summary and outlook for the future.

\section{The Supersymmetric Framework.}\label{sec:framework}

\subsection{Particle Content}

We begin by considering the field content of the minimal supersymmetric
extension of the SM.
Each chiral fermion $f_{L,R}$ in the SM has a spin zero supersymmetric
partner, the sfermion $\tf_{L,R}$ ($f=q,\ell$) with the same gauge quantum
numbers, so that the number of
bosonic and fermionic degrees of freedom are the same. Note that this means
that for each
massive Dirac fermion there are {\it two} distinct scalar partners ($\tf_L$ and
$\tf_R$), each represented by a
complex scalar field. The chiral fermion field with the corresponding scalar
field
and the auxiliary field necessary to linearly realize the SUSY
algebra~\cite{Bagger} together
constitute a chiral matter supermultiplet, in much the same way that the proton
and the neutron
together constitute the isospin doublet. The complex Higgs bosons together with
their
spin $\frac{1}{2}$ partners, the Higgsinos (and the associated auxiliary field)
comprise
the Higgs chiral supermultiplets. Finally,
the partner of a massless spin-1 real gauge field
is a spin $\frac{1}{2}$ Majorana~\footnote{A Majorana spinor $\psi$ is one
whose charge conjugate
$\psi^c=\psi$. Notice that this means that $\psi_L$  and $\psi_R^*$ are
linearly related.}
gaugino which together with a real auxiliary field constitute the gauge
supermultiplet.~\footnote{We
choose the Wess-Zumino gauge discussed by Bagger. Again, note the equality of
the physical
bosonic and fermionic degrees of freedom.}

All SUSY models must, therefore, include quark and lepton chiral
superfields (a superfield~\cite{Bagger} is simply a supermultiplet
whose components are fields, in exactly the same way that the Yang-Mills
field is a gauge multiplet whose components are
(gauge) fields) with gauge quantum
numbers corresponding to those of the correponding quark or lepton.
For instance, there is an SU(3)
triplet, SU(2) singlet superfield $\hat{U}_R$ with $Y=\frac{4}{3}$ (the
subscript R reminds us
of the chirality of the corresponding SM fermion) {\it etc.} Since the
superpotential is required to be a
function of only left (or equivalently, only right) chiral superfields, we will
instead of $\hat{U}_R$ work
with the anti-quark SU(2) singlet superfield $\hat{U}^c$ (henceforth, we drop
the redundant chirality
subscript since all fields are left-chiral), and likewise introduce weak
isosinglet superfields $\hat{D}^c$
and $\hat{E}^c$ in addition to the quark and lepton doublets $\hat{Q}$ and
$\hat{L}$, respectively. As
in the SM, the replication of generations must be put in by hand.
We must also include gauge super-multiplets which
transform according to the adjoint representation of the gauge group. Finally,
we have to
introduce at least
two different Higgs supermultiplets $\hat{h}$ and $\hat{h'}$ to give masses to
the up- and down- type
fermions, respectively.~\footnote{The Yukawa interactions come from
superpotential interactions. Since
the superpotential is required to contain only left-chiral superfields, we
cannot use the
the Higgs field and its conjugate (which will have opposite chirality)
simultaneously in the superpotential.
This is unlike the situation in
the SM where the sole Higgs field $\phi$ and its conjugate $\phi^*$ can be
simultaneously introduced
into the Lagrangian.} We will use the notation of the Korea
Lectures~\cite{Korea},
and require $\hat{h}$ ($\hat{h'}$) transform as
the $2$ ($2^*$) representation of $SU(2)$. The Minimal Supersymmetric Model
(MSSM) contains the smallest
number of new fields --- the matter and gauge multiplets along with exactly two
Higgs doublets.

\subsection{Supersymmetric Interactions}

Bagger has already discussed how to construct supersymmetric Lagrangians in his
lectures.~\cite{Bagger} We will not repeat this discussion here, but only
recapitulate
the necessary results. The Lagrangian for global supersymmetry, after
elimination of auxiliary fields, can be written
in four-component spinor notation as,~\cite{Korea}

\begin{eqnarray}
\cal{L} & =& \sum_{i} (D_{\mu}S_i)^{\dagger}(D^{\mu}S_i) +
\frac{i}{2}\sum_{i}\bar{\psi_i}\rlap /D \psi_i \nonumber\\
  &  & -\frac{1}{4}\sum_A F_{\mu \nu A}F^{\mu \nu}_A + \frac{i}{2}\sum_A
\bar{\lambda}_A\rlap /D \lambda_A\nonumber\\
  &  & -\sqrt{2}\sum_i\left[S_i^\dagger(t_A)\bar{\psi_i} \frac{1-\gamma_5}{2}
\lambda_A + h. c.\right]\nonumber\\
  &  & -\frac{1}{2}\sum_A\left[\sum_i S_i^\dagger t_A S_i +
\xi_A\right]^2\nonumber\\
  &  & -\sum_i \left|\frac{\partial f}{\partial S_i}\right|^2\nonumber\\
  &  & -\frac{1}{2}\sum_{i,j}\left\{\bar{\psi_i}\left[\frac{1-\gamma_5}{2}
\right]\frac{\partial^2 f}{\partial S_i
 \partial S_j}\psi_j +h.c.\right\}
\label{eq:lagrangian}
\end{eqnarray}

Here, $S_i$ ($\psi_i$) denotes the scalar (Majorana fermion~\footnote{The
four-component spinor is
defined~\cite{Korea} by choosing the
right chiral component such that the spinor is Majorana.}) component of the
{\it i}th chiral superfield,
$F_{\mu \nu A}$ is the Yang-Mills gauge field, and $\lambda_A$ is the Majorana
gaugino superpartner of the corresponding
gauge boson and $\xi_A$ are constants which can be non-zero only for U(1)
factors of the gauge group.~\cite{FI}
In anticipation of simple grand unification, we will set these to zero. The
function $f$ in the last
two lines of Eq.~(\ref{eq:lagrangian}) is the superpotential which is a
function of chiral superfields. By
$\frac{\partial f}{\partial S_i}$ (and other derivatives), we mean
differentiate with respect to $\hat{S_i}$ and
then set $\hat{S_i}=S_i$.

We note the following:
\begin{enumerate}

\item The first two lines are the gauge invariant kinetic energies for the
components of
the chiral and gauge superfields. The derivatives that appear are gauge
covariant derivatives
appropriate to the particular representation in which the field belongs. For
example, if we are
talking about SUSY QCD, for quark fields in the first line of
Eq.~(\ref{eq:lagrangian}) the covariant
derivative contains triplet SU(3) matrices, whereas the covariant derivative
acting on the gauginos
in the following line will contain octet matrices. As stressed above,
these terms completely determine how particles interact with
gauge bosons.

\item The next line describes the interactions of gauginos  with matter and
Higgs multiplets.
Notice that these interactions are also determined by the gauge couplings. Here
$t_A$ is
the appropriate dimensional matrix represention of the group generators times
the gauge coupling constant.
Matrix multiplication is implied. To see that these terms are gauge invariant,
recall that $\psi_{iR}$ which is
fixed by the Majorana condition, transforms according to the conjugate
representation to $\psi_{iL}$.

\item Line four describes the quartic couplings of scalar matter. Notice that
these are determined by
the gauge interactions. The interactions on this line are referred to as
$D$-terms.

\item Finally, the last two lines in Eq.~(\ref{eq:lagrangian}) describe the
non-gauge superpotential interactions of matter
fields and lead to the Yukawa interactions responsible for matter fermion
masses in the SM. Since these interactions do not
involve any spacetime derivatives, choosing the superpotential to be a globally
gauge invariant function of superfields is
sufficient to guarantee the gauge invariance of the Lagrangian. For a
renormalizable theory, the superpotential
must be a polynomial of degree $\leq  3$.

\end{enumerate}

Assuming the minimal field content discussed above
and neglecting intergenerational mixing, the most general
SU(3)$\times$SU(2)$\times$U(1) invariant superpotential
can be written as,

\begin{equation}
f = f_1 + g_1 + g_2,
\label{eq:sup}
\end{equation}

with
\begin{eqnarray}
f_1 &=& \mu(\hat{h}^0\hat{h'}^0 + \hat{h}^+\hat{h'}^-) +
f_u(\hat{u}\hat{h}^0 -\hat{d}\hat{h}^+)\hat{U}^c\nonumber \\
    & & f_d(\hat{u}\hat{h'}^- + \hat{d}\hat{h'}^0)\hat{D}^c
+ f_e(\hat{\nu}\hat{h'}^- + \hat{e}\hat{h'}^0)\hat{E^c} + \ldots,
\label{eq:supR}
\end{eqnarray}

\begin{equation}
g_1 = \sum_{i,j,k}\left[\lambda_{ijk}\hat{L_i}\hat{L_j}\hat{E_k}^c +
\lambda^{'}_{ijk}\hat{L_i}\hat{Q_j}\hat{D_k}^c\right],
\label{eq:supL}
\end{equation}

and,

\begin{equation}
g_2 = \sum_{i,j,k} \lambda^{''}_{ijk}\hat{U_i}^c\hat{D_j}^c\hat{D_k}^c.
\label{eq:supB}
\end{equation}
Other models may be constructed by introducing additional fields into the
superpotential.

In Eq.~(\ref{eq:supR}), $\hat{u}$ and $\hat{d}$ denote the doublet quark
superfields. A similar
notation is used for leptons. The minus sign in the second term is because it
is the anti-symmetric
combination of two doublets that forms an SU(2) singlet. Since $\hat{h'}$ is
defined
to transform according to the $2^*$ representation, the symmetric combination
appears in other terms.
Also, $f_u$, $f_d$ and $f_e$ are the coupling constants for the Yukawa
interactions that give rise to
first generation quark and lepton masses.
The ellipses denote similar terms for other generations. In the
Eq.~(\ref{eq:supL}) and (\ref{eq:supB}),
 $i,j$ and $k$ denote generation indices, while the $\lambda$'s are coupling
constants.~\footnote{We have omitted
a bilinear term in $g_1$. This term can be rotated away in the supersymmetry
limit.} We have, for brevity, not
expanded out the gauge invariant product of doublets in Eq.~(\ref{eq:supL}).
The Lagrangian interactions
can now be obtained by substituting the superpotential (\ref{eq:sup}) into
Eq.~(\ref{eq:lagrangian}).
It is easy to check that the terms obtained from
$g_1$ and $g_2$ lead to the violation of lepton and baryon number conservation,
respectively. This can also be seen directly from the superpotential: for
instance, with the usual assignment of
lepton number of one unit to $\hat{L}$ and $\hat{E}$ (so that $f_1$ remains
invariant),
$g_1$ clearly is not globally invariant under the corresponding U(1)
transformations.

This situation is
quite different from the SM where the gauge invariance of the Lagrangian
guarantees the absence of renormalizable
baryon or lepton number violating interactions. It is the presence of scalar
baryon and lepton superpartners
that now allow for renormalizable baryon and lepton number violating
fermion-fermion-scalar vertices.
The simultaneous presence of all such terms with large couplings would lead to
proton decay at the weak
interaction rate if, as Bagger has explained, the superpartners have masses
below the TeV scale.
This would, of course, be a phenomenological disaster. Unlike as in the SM, an
additional global symmetry
needs to be put in by hand (or some dimensionless couplings need to be chosen
to be tiny) to prevent
this.
One possible way~\footnote{It has, however,
recently been argued~\cite{SHER}
that certain products of B- and L-violating interactions, even if
simultaneously present, are only weakly
constrained.} to guarantee proton stability
is to assume that at least one (or both) of baryon or lepton number is
conserved (and in the case
of baryon number violation, also that the lightest SUSY fermion is heavier than
the proton). Within the
 MSSM it is assumed that both $g_1$ and $g_2$ vanish; {\it i.e.} the model is
minimal
in that it not only contains the fewest new particles, but also
the fewest number of interactions necessary to be phenomenologically viable.
It is easy to check that the interactions then multiplicatively
conserve a new quantum number called $R$-parity which is defined
to be +1 for SM particles such as quarks, leptons and gauge and Higgs bosons,
and -1 for their supersymmetric
partners.~\footnote{Notice that $R$-parity is automatically conserved by the
interactions of gauge bosons
and gauginos on the first four lines of Eq.~(\ref{eq:lagrangian}). Whether or
not it is a good symmetry
then depends on the choice of superpotential. Spontaneous $R$-violation via a
vacuum expectation value
of a doublet sneutrino is excluded by the measurement of the $Z$ width at LEP,
as we will discuss later.}
It is also possible to construct phenomenologically viable models that include
$g_1$ or $g_2$ terms in the
superpotential and so violate $R$-parity conservation. We will return to such
non-minimal models at
the end of these lectures but will focus, for now, on the MSSM.

\subsection{Supersymmetry Breaking}

The interactions defined by the Lagrangian (\ref{eq:lagrangian}) are exactly
supersymmetric, and so, cannot be the
whole story. The physics of supersymmetry breaking is, however, not understood
so that the best that we can
hope for at present is a parametrization of SUSY-breaking effects. The guiding
principle, as we have
already noted, is that the SUSY breaking terms should not destabilize scalar
masses by
reintroducing the quadratic divergences that SUSY was introduced to eliminate
in the first place.
Girardello and Grisaru~\cite{GG} have classified all renormalizable soft SUSY
breaking operators. For our purposes, it is
sufficient to know that these consist of,
\begin{itemize}

\item explicit masses for the scalar members of chiral multiplets; {\it i.e.}
squarks, sleptons
and Higgs bosons,

\item explicit masses $\mu_1$, $\mu_2$ and $\mu_3$ for the U(1), SU(2) and
SU(3) gauginos,

\item new super-renormalizable scalar interactions: for each trilinear
(bilinear) term in the superpotential of the form
$C_{ijk}\hat{S_i}\hat{S_j}\hat{S_k}$ ($C_{ij}\hat{S_i}\hat{S_j}$), we
can introduce a soft supersymmetry breaking scalar interaction
$A_{ijk}C_{ijk}{S_i}{S_j}{S_k}$ ($B_{ij}C_{ij}{S_i}{S_j}$)
where the $A$'s and $B$'s are constants. These terms are often referred to as
$A$- and $B$-terms.

\end{itemize}
The scalar and gaugino masses obviously serve to break the undesired degeneracy
between the masses of sparticles
and particles. We will see later that the explicit trilinear scalar
interactions mainly affect the phenomenology of third
generation sfermions.

\section{The Minimal Supersymmetric Model}\label{sec:mssm}

The MSSM is the simplest supersymmetric extension of the SM in that it contains
the fewest number
of fields and superpotential interactions. Here, we identify the particles,
{\it i.e.} the mass eigenstates
of the model, and also summarize the model parameters that have been
introduced.

\subsection{Mass Eigenstates and their Interactions}

{\it SUSY Scalars:} The scalar partners $\tf_L$ and $\tf_R$ have the same
electric charge and colour,
and so can mix if SU(2)$\times$U(1) is broken. It is simple to check that the
gauge interactions
conserve chiral flavour in that they couple only left (right) multiplets with
one another, {\it i.e.}
$\tf_L$ couples only to $\tf_L$ ($f_L$) via gauge boson (gaugino) interactions.
Unless this ``extended
chiral symmetry'' is broken, there can be no $\tf_L-\tf_R$ mixing. This
symmmetry is, however, explicitly
broken by the Yukawa interactions in the superpotential.~\footnote{Without the
assumption of
$R$-parity consevation there would also be mixing between the $\hat{h}$ and
$\hat{L}$
supermultiplets. Such a mixing which is absent in the MSSM can have significant
phenomenological impact.} We thus conclude that $\tf_L-\tf_R$ mixing
is proportional to the corresponding Yukawa coupling and hence to the
corresponding {\it fermion} mass.
This mixing is generally negligible except for the case of top squarks where it
plays a very
important role. We will, therefore, neglect this intra-generational mixing for
the first five
squark flavours, and for simplicity, also any inter-generational mixing.

{\it SUSY Fermions:} The gauginos and Higgsinos are the only spin-$\frac{1}{2}$
fermions.
Of these, the gluinos being the only colour octet fermions, remain unmixed  and
have a mass $m_{\tg} = |\mu_3|$.

Electroweak gauginos
and Higgsinos of the same charge can mix, once electroweak gauge invariance is
broken.
The mass matrices can be readily worked out using Eq.~(\ref{eq:lagrangian}),
(\ref{eq:sup})
and  (\ref{eq:supR}). These matrices for both the charged and neutral
electroweak
gaugino sector are explicitly given elsewhere~\cite{Korea} and will not be
rewritten
here. Note the translation, $\mu \equiv -2m_1$.
The mass eigenstates can be obtained by diagonalizing these
matrices.~\footnote{If
an eigenvalue of the mass matrix for any state $\psi$
turns out to be negative, one can always define a new
spinor $\psi^{'}=\gamma_5\psi$ which will have a positive mass. For
neutralinos, $\psi^{'}$
should be defined with an additional factor $i$ to preserve its Majorana nature
under the $\gamma_5$ transformation.}
In the MSSM
the charged Dirac Higgsino (composed of the charged components of the doublets
$\tilde{h}$ and $\tilde{h^{'}}$) and the charged gaugino (the partner of the
charged $W$ boson)
mix to form two Dirac charginos, $\tw_1$ and $\tw_2$, while the two neutral
Higgsinos
and the neutral SU(2) and U(1) gauginos mix to form four Majorana neutralinos
$\tz_1 \ldots \tz_4$,
in order of increasing mass. In general, the mixing patterns are complex and
depend
on several parameters: $\mu$, $\mu_{1,2}$ and $\tan\beta \equiv \frac{v}{v'}$,
the ratio
of the vacuum expectation values of the two Higgs fields introduced above.
If either $|\mu|$ or $|\mu_1|$ and $|\mu_2|$
are very large compared to $M_W$, the mixing becomes small. For $|\mu| >> M_W,
|\mu_{1,2}|$, the lighter
chargino is essentially a gaugino while the heavier one is a Higgsino with mass
$|\mu|$; also,
the two lighter neutralinos are gaugino-like while $\tz_{3,4}$ are dominantly
Higgsinos
with mass $\sim |\mu|$. If instead, the gaugino masses are very large, it is
the heavier
chargino and neutralinos that become gaugino-like.

Without further assumptions, the three gaugino masses are independent
parameters. It is, however,
traditional to assume that there is an underlying grand unification, and that
these masses
derive from a common gaugino mass parameter defined at the unification scale.
The differences between the various gaugino masses then come from the fact that
they
have different interactions, and so, undergo different renormalization when
these are
evolved down from the GUT scale to the weak scale. The gaugino masses are then
related by,

\begin{equation}
\frac{3\mu_1}{5\alpha_1} = \frac{\mu_2}{\alpha_2} = \frac{\mu_3}{\alpha_3}.
\label{eq:gaugino}
\end{equation}
Here the $\alpha_i$ are the fine structure constants for the different factors
of the gauge group.
With this GUT assumption, $\tw_1$  and $\tz_{1,2}$ will be substantially
lighter than gluinos.
It is for this reason that future $e^+e^-$ colliders operating at
$\sqrt{s} \simeq 500$-1000~GeV
are expected to be competitive
with hadron supercolliders such as the LHC which has much higher energy. We
also mention that
for not too small values of $|\mu|$,
the lightest neutralino tends to be dominantly the hypercharge gaugino.

{\it The Electroweak Symmetry Breaking Sector:}
Although this is not in the mainstream of what we will discuss, we should
mention
that because there are two doublets in the MSSM, after the Higgs mechanism
there are
five physical spin zero Higgs sector particles left over in the spectrum.
Assuming that there are
no CP violating interactions in this sector, these are two neutral CP even
eigenstates ($H_{\ell}$ and
$H_h$) which
behave as scalars as far as their couplings to fermions go (the subscripts
$\ell$ and $h$ denote
light and heavy), a neutral ``pseudoscalar'' CP odd particle $H_p$, and a pair
of charged particles $H^{\pm}$.

The Higgs boson sector~\cite{HHG} of the MSSM is greatly restricted by SUSY. In
addition to $\tan\beta$ which also
enters the gaugino-Higgsino sector, the tree level Higgs sector is fixed by
just one additional
parameter which my be taken to be $m_{H_p}$.
In particular, the Higgs
quartic self-couplings are all given by those on line four of
Eq.~(\ref{eq:lagrangian})
and so are fixed to be \cal{O}($g^2$). This leads to the  famous (tree-level)
bound,
$m_{H_{\ell}} < \min [M_Z, m_{H_p}]|\cos2\beta|$. This receives
important corrections from $t$ and $\tt$
loops because of the rather large value of the top Yukawa coupling and the
bound is weakened~\cite{OK}
to about 120-130~GeV depending on the value of $m_t$. Thus, in contrast to
early expectations,
$H_{\ell}$ may well escape detection at LEP2. It is worth mentioning that if we
assume
that all couplings remain perturbative up to the GUT scale, then the mass of
the lightest
Higgs boson is bounded by 145-150~GeV in {\it any} weak scale SUSY
model.~\cite{KOLD} The physics
behind this is the same as that behind the
bound~\cite{CAB} $m_{H_{SM}} \alt 200$~GeV on the mass of the SM Higgs boson,
obtained under the assumption
that the Higgs self-coupling not blow up below the GUT scale; the numerical
difference
between the bounds comes from the difference in the evolution of the running
couplings
in SUSY and the SM. An $e^+e^-$ collider operating at a centre of mass energy
$\sim 300$~GeV
would thus be certain~\cite{OKAD} to find a Higgs boson if these arguments are
valid.

\subsection {MSSM Model Parameters: A Recapitulation}

For the convenience of the reader and for subsequent developments, we first
summarize the
parameters of the MSSM. In addition to the SM parameters, the MSSM parameters
include,

\begin{itemize}
\item $\tan\beta$ and the superpotential parameter $\mu$,
\item soft breaking masses for each of the three gauginos: these are given in
terms of a single
parameter if we assume the gaugino mass unification
condition~(\ref{eq:gaugino}).
\item There is an independent soft SUSY breaking
scalar mass for each SM; {\it i.e.} SU(3)$\times$SU(2)$\times$U(1), matter
multiplet. There are thus six slepton masses and
nine squark masses for the three families, even if
mixing between the generations is ignored.
\item Again without inter-generational mixing, there are nine $A$-parameters,
and
a $B$-parameter for the one bilinear term in the MSSM superpotential $f_1$.
\item Finally, there is the one additional parameter (chosen to be $m_{H_p}$)
that determines the tree-level
Higgs boson sector.
\end{itemize}

We see that without assuming anything more than SU(3)$\times$SU(2)$\times$U(1)
invariance, the model
contains an unmanageably large number of parameters. Assuming grand unification
ameliorates the situation
to some extent: there are then only two scalar masses per generation of
sfermions in SU(5) and only
one gaugino mass parameter, but the parameter space is still too large for the
phenomenology to be tractable.
Inspired by supergravity model studies, many early phenomenological studies
assumed that all
squarks (sleptons were either assumed to be degenerate with squarks, or to have
masses
related to $m_{\tq}$) were degenerate except for $D$-term splitting. They also
incorporated the GUT assumption for gaugino masses. In this case the masses and
couplings
of all sparticles were determined in terms of relatively few SUSY parameters
which were frequently
taken to be,

\begin{equation}
m_{\tg}, m_{\tq}, m_{\tell}, \mu, \tan\beta, A_t, m_{H_p}.
\label{eq:mssm}
\end{equation}
The parameter $A_t$ mainly affects top squark phenomenology, and so, was
frequently irrelevant.
Other $A$-terms, being proportional to the light fermion masses, are
negligible.

In view of the fact that additional assumptions are necessary, and further,
that
assumptions based on supergravity models are incorporated
into phenomenological analyses, it seems reasonable to explore the implications
of these
models more seriously. Toward this end, we describe the underlying framework
in the following section.

\section{Minimal Supergravity Models}\label{sec:sugra}

When supersymmetry is promoted to a local symmetry, additional fields have to
be introduced.
The resulting theory~\cite{GRAV} which includes gravitation is known as
supergravity (SUGRA). It is not our purpose
here to study SUGRA models in any detail. In fact, local supersymmetry will not
play
any direct role in our later considerations. The purpose of this discussion is
merely to maintain
continuity of development, and also to provide motivation for an economic and
elegant framework
that has recently become very popular for phenomenological
analysis.~\footnote{Although this has been discussed
by Bagger~\cite{Bagger} it seems necessary to include discussion of this
important topic for completeness.}

It was recognised rather early that it is very difficult to construct
globally supersymmetric models where SUSY is spontaneously broken
at the weak scale. This led to the development of geometric hierarchy models
where SUSY is broken in a ``hidden'' sector at a scale $\mu_s >> M_W$. This
sector
is assumed to interact with ordinary particles and their superpartners (the
``observable''
sector) only via exchange of superheavy particles $X$. This then suppresses the
couplings
of the Goldstone fermion (which resides in the hidden sector) to  the
observable sector: as
a result, the effective mass gap in the observable sector is $\mu \sim
\frac{\mu_{s}^{2}}{M_X}$
which can easily be $\alt 1$~TeV even if $\mu_s$ is much larger.

An especially attractive realization of this idea stems from the assumption
that the hidden
and observable sectors interact only gravitationally, so that the scale $M_X$
is $\sim M_{Planck}$.
This led to the development of SUGRA GUT models of particle physics. Because
supergravity is not
a renormalizable theory, we should look upon the resulting Lagrangian, with
heavy degrees of freedom
integrated out, as an effective theory valid below some ultra-high scale $M_X$
around $M_{GUT}$ or $M_{Planck}$,
in the same way that chiral
dynamics describes interactions of pions below the scale of chiral symmetry
breaking.
Remarkably, this Lagrangian turns out to be just the same as that of
a globally supersymmetric SU(3)$\times$SU(2)$\times$U(1)
model, together with soft SUSY breaking masses and $A$- and $B$-parameters
of $O(M_{Weak})$.

The economy of the minimal supergravity GUT framework~\footnote{Here the term
minimal refers to the canonical
choice of kinetic energy terms for matter and gauge fields. Since supergravity
is a non-renormalizable
theory, in principle, these terms can arise from higher dimensional operators.}
stems from the fact
that because of the assumed symmetries, various soft SUSY breaking parameters
become related
independent of the details of the hidden sector and the low energy effective
Lagrangian
can be parametrized in terms of just a few parameters. For instance, since the
chiral multiplets
are universally coupled to the hidden sector (via gravitational interactions),
they all acquire
the same soft SUSY breaking scalar mass $m_0$. Likewise, there is a universal
$A$-parameter,
common to all  trilinear interactions. The GUT assumption, of course, implies
that the soft
SUSY breaking gaugino masses are related as in Eq.~(\ref{eq:gaugino}).
It should be emphasized that the universality of the scalar
masses does not imply that the physical scalar masses of all sfermions are the
same. The point is
that the parameters in
the Lagrangian obtained by integrating out heavy fields
should be regarded as renormalized at the high scale $M_X$ at which these
symmetries are unbroken.
If we use this Lagrangian to compute processes at the 100~GeV energy scale
relevant
for phenomenology, large logarithms \cal{O}($\ln\frac{M_X}{M_W}$) due to the
disparity between
the two scales invalidate the perturbation
expansion. These logarithms can be straightforwardly summed by evolving the
Lagrangian
parameters down to the weak scale.
This is most conveniently done~\cite{INOUE} using renormalization group
equations (RGE).

The renormalization group evolution leads to an interesting pattern of
sparticle masses, evaluated
at the weak scale.~\footnote{These running masses evaluated at
the sparticle mass, or more crudely, at a scale $\sim M_Z$, are not identical
to, but are frequently close
to the physical masses which are given by the pole of the renormalized
propagator~\cite{POLE}.}
For example,
gauge boson-gaugino loops result in increased sfermion masses as we evolve
these
down from $M_X$ to $M_W$ while superpotential Yukawa couplings (which are
negligible for the
two lightest generations) have just the opposite effect.
Since squarks have strong interactions in addition to the electroweak
interactions
common to all sfermions, the weak scale squark masses are larger than those of
sleptons. Neglecting
Yukawa couplings, we have to a good
approximation,

\begin{eqnarray}
m_{\tq}^2 & = & m_0^2 + m_q^2 + (5-6)\mhf^2 + D-terms, \nonumber \\
m_{\tell}^2 & = & m_0^2 + m_{\ell}^2 + (0.15-0.5)\mhf^2 + D-terms.
\label{eq:sfermions}
\end{eqnarray}

In Eq.~(\ref{eq:sfermions}), $\mhf$ is the common gaugino mass at the scale
$m_X$. Notice that
squarks and sleptons within the same SU(2) doublets are split by the $D$-terms,
once electroweak
symmetry is broken. In contrast, various flavours of left- (and separately,
right-) type squarks of the
first two generations are essentially
degenerate, consistent~\cite{FCNC}
with flavour changing neutral current (FCNC) constraints in the $K$-meson
sector.~\footnote{This is a non-trivial
observation since alternative mechanisms to suppress FCNC based on different
symmetry considerations have been
proposed.~\cite{NS}}
The $D$- terms, which are  typically $\leq \frac{1}{2}M_Z^2$, are generally
not important when sfermions are heavy. The difference in the coefficients of
the $\mhf$ terms
reflects the difference between the strong and electroweak interactions alluded
to above. Although
we have not shown this explicitly, $\tell_R$ which has only hypercharge
interactions tends to be
lighter than $\tell_L$ as well as $\tnu_L$ unless D-term effects are
significant.
Since $m_{\tg} = (2.5-3)\mhf$, it is easy to see that squark and slepton masses
are related by,

\begin{equation}
m_{\tq}^2 = m_{\tell}^2 + (0.7-0.8){m_{\tg}}^2.
\label{eq:squark}
\end{equation}
Here, $m_{\tq}^2$ and $m_{\tell}^2$ are the squared masses averaged over the
squarks or sleptons of the first (or second) generation. In the second term,
the unification of gaugino masses
has been assumed. Since experimental data, as we will see, requires squarks to
be heavier than 150-200~GeV,
it immediately follows that the first two generations of squarks are
approximately degenerate.

The Yukawa couplings of the top family are certainly not negligible. For very
large
values of
$\tan\beta \sim \frac{m_t}{m_b}$ bottom Yukawa couplings are also important. As
mentioned above,
these Yukawa interactions tend to reduce the scalar masses at the weak scale.
These corrections
can overcome the additional $m_t^2$ in Eq.~(\ref{eq:sfermions}), so that
$\tt_L$ and $\tt_R$
tend to be significantly lighter than other squarks (of course, by SU(2)
invariance, the soft-breaking
mass for $\tb_L$ is the same as that for $\tt_L$). In fact, we can say more:
because $\tt_R$ receives
top quark Yukawa corrections from both charged and neutral Higgs loops in
contrast to $\tt_L$ which
gets corrections just from the neutral Higgs, its squared mass is reduced by
(approximately) twice as much
as that of $\tt_L$.
Moreover, as we have already seen, these same Yukawa
interactions lead to $\tt_L-\tt_R$ mixing, which further depresses the mass of
the lighter of the two
$t$-squarks (sometimes referred to as the stop)
which we will denote by $\tt_1$. In fact, care must be exercised in the choice
of input parameters:
otherwise $m_{\tt_1}^2$ is driven negative, leading to the spontaneous
breakdown of
electric charge and colour.

The real beauty and economy of this picture comes from the fact that these same
Yukawa radiative corrections
drive~\cite{RAD}
electroweak symmetry breaking. Since the Higgs bosons are part of chiral
supermultiplets, they also have a common
mass $m_0$ at the scale $M_X$
and undergo similar renormalization as doublet sleptons
due to gauge interactions;
{\it i.e.} these positive contributions are not very large. The squared
mass $m_h^2$ of the Higgs boson doublet $h$ which
couples to the top family, however, receives large negative contributions
(thrice those of the $\tt_L$ squark since
there are three different colours running in the loop) from Yukawa
interactions, and so can become negative, leading
to the correct pattern of gauge symmetry breaking. Furthermore, because $f_t >
f_b$, $\tan\beta > 1$.
While this mechanism is indeed very pretty, it should probably
not be regarded as an explanation of the observed scale of
spontaneous symmetry breakdown since it requires that $m_0$, the scalar
mass at the very large scale $M_X$ be chosen to be $\leq 1$~TeV: in other
words, the small dimensionless ratio
$\frac{m_0}{M_X}$ remains unexplained.

Let us compare the model parameters with our list (\ref{eq:mssm}) for the MSSM.
Within SUGRA
GUTS, we start with GUT scale parameters, $m_0$, $\mhf$, $A_0$, $B_0$ and
$\mu_0$. The weak scale
parameter $\mu$ (actually, $\mu^2$) is adjusted to give the experimental value
of $M_Z$.
It is convenient to eliminate $B_0$ in favour of $\tan\beta$ so that the model
is completely specified
by just four parameter set (with a sign ambiguity for $\mu$),

\begin{equation}
m_0, \mhf, \tan\beta, A_0, sgn(\mu),
\label{eq:sugra}
\end{equation}
without the need of additional {\it ad hoc} assumptions as in the MSSM.
Comparing with the MSSM parameter
set~(\ref{eq:mssm}) we see that $\mu$ and $m_{H_p}$ are no longer free
parameters.

SUGRA models lead to a rather
characteristic pattern of sparticle masses~\cite{SPECTRA} and mixings. We have
already seen that the first
two generations of squarks are approximately degenerate, while the lighter of
the $t$-squarks,
and also $\tb_L$ can be substantially lighter. Also, from Eq.~(\ref{eq:squark})
it follows that
sleptons may be significantly lighter than the first two generations of
squarks if $m_{\tg} \simeq m_{\tq}$, and have
comparable masses if squarks are significantly heavier than gluinos. We also
see that
gluinos can never be much heavier than
squarks.~\footnote{Ellwanger~\cite{Ellwanger} has shown that this
is a very general result not special to supergravity. It follows from the
requirement that
$m_{\tq}^2$ not be driven to negative values below the unification scale,
assuming only that there are
no new large Yukawa interactions.} Furthermore, because the top quark is very
massive, the value of $|\mu|$ obtained from the radiative
symmetry breaking constraint generally tends to be much larger than the
electroweak gaugino masses,
so that the lighter (heavier) charginos and neutralinos tend to be gaugino-like
(Higgsino-like).

It should be kept in mind that while the minimal SUGRA framework provides a
very attractive and economic
picture, it hinges upon untested assumptions of symmetries about the physics at
very high energies. It could be
that the GUT assumption is incorrect though this would then require the
unification implied~\cite{Paul} by the
observed values of gauge couplings at LEP to be purely fortituous. It could be
that the assumption
of universal scalar masses (or $A$-parameters) is wrong. Recall that although
we used supergravity couplings
between the hidden and observable sectors to argue for this, the common scalar
mass was a consequence
of the assumed universality of the (gravitational) couplings between the hidden
and observable sectors.
In other words, the universality of scalar masses is really a result of an
assumed
global U(N) symmetry of the Lagrangian for transformations amongst the N chiral
supermultiplets, an assumption
which is, perhaps, reasonable as long as we are near the Planck scale where
gravitation presumably
dominates GUT gauge
interactions. Non-universal masses could result if this U(N) is broken as
Bagger has illustrated in his
lectures by the explicit introduction of non-renormalizable terms in the
superpotential.
We should also remember that in the absence of a theory about physics at the
high scale,
we do not have a really good principle for choosing the scale $M_X$ at which
the scalar masses are universal. In practice, most phenomenological
calculations set this to be the scale of GUT symmetry
breaking where the gauge couplings unify. If, instead, this scale were closer
to $M_{Planck}$ the evolution between these scales~\cite{PP} could result
in non-universal scalar masses at $M_{GUT}$: this could have significant
impact, particularly on the condition of
electroweak symmetry breaking.

Despite these shortcomings, this framework at the very least should be expected
to provide a useful guide to
our thinking about supersymmetry phenomenology. In spite of the fact that it is
theoretically rather constrained,
it is consistent with all experimental and even
cosmological constraints and even, as we will see, contains a candidate for
galactic and cosmological dark matter.
But it should be kept in mind that some of the underlying assumptions may prove
to be incorrect. For this  reason,
one should always be careful to test the sensitivity of the phenomenological
predictions to the various
assumptions, especially when considering the design of future experiments. It
is, nevertheless, worth emphasizing that we now
have a reasonably flexible yet tractable framework whose underlying
assumptions, as we will see, can be
subject to direct tests at future colliders.

\section{Decays of Supersymmetric Particles}\label{sec:decays}

Before we can discuss signatures via which sparticle production might be
detectable at colliders, we need
to understand how sparticles decay. The conservation of $R$-parity implies that
sparticles can only decay
into other sparticles, until the decay cascade terminates in the lightest
supersymmetric particle (LSP) which
is absolutely stable. There are strong limits~\cite{Norman}
 on the existence of stable or even very long-lived ($\tau >$ age
of the universe) coloured or charged sparticles. Such sparticles, which would
have been abundantly produced in the
Big Bang, would bind to ordinary particles to form
exotic atoms or nuclei~\cite{Wolfram}. For masses up to 1~TeV, their
expected abundances are in the range $\agt 10^{-10}$, whereas we know
experimentally that these abundances
are $<$~ \cal{O}($10^{-12}-10^{-29}$) depending on the new particle mass and
also on the
element whose exotic isotope is being searched for.

Within the MSSM, the null result of these searches is taken to imply that the
LSP must be a weakly interacting
neutral particle; {\it i.e.} it must be either the lightest neutralino $\tz_1$,
or one of the sneutrinos. We
will see later that sneutrinos are excluded as the LSP if we also require that
they make up galactic dark matter.
Within the SUGRA
framework, the LSP could also be the gravitino --- the SUSY partner of the
graviton. Unless it is extremely light,
it couples to other particles with gravitational strength couplings, so that it
is effectively decoupled for
the purposes of collider phenomenology: then, the next lightest SUSY particle,
which will
decay outside the detector, plays the role of the LSP.~\footnote{As long as the
next lightest
sparticle is neutral, SUSY phenomenology at colliders is essentially unaltered.
The late decay
of  this effective LSP can potentially spoil the successful predictions of Big
Bang nucleosynthesis
as discussed by Moroi.~\cite{MOROI}}
In this case (or if $R$-parity is not conserved),
however, the ``effective'' (or actual) LSP
may even be charged or coloured. Throughout most of these lectures, we will
assume that $\tz_1$ is the LSP.

We note here that regardless of details the neutral LSP's which are produced at
the termination
of the SUSY decay cascade behave like stable, heavy neutrinos in the
experimental apparatus in that
they escape without depositing any energy. Thus apparent missing energy
($E\llap/$) and an imbalance
of transverse momentum ($\pslt$) are generally regarded as canonical signatures
of supersymmetry.

\subsection{Sfermion Decays}

We have seen in Sec. \ref{sec:framework} that gauginos and Higgsinos couple
sfermions to fermions.
Since we have
also assumed that $\tz_1$ is the LSP, the decay $\tf_{L,R} \to f\tz_1$ ($f \not
= t$)
is always allowed. Depending on sparticle masses, the decays
\begin{equation}
\tf_{L,R} \to f\tz_i, \tf_{L} \to f'\tw_i
\label{eq:sfermiondk}
\end{equation}
to other neutralinos or to charginos may also be allowed. The chargino decay
modes of $\tf_R$ only proceed
via Yukawa interactions, and so are negligible for all but $t$-squarks. Unlike
sleptons, squarks also have
strong interactions, and so can also decay into gluinos via,
\begin{equation}
\tq_{L,R} \to q\tg,
\label{eq:squarkdk}
\end{equation}
if $m_{\tq}>m_{\tg}$. Unless suppressed by phase space, the gluino decay mode
of squarks dominates,
so that squark signatures are then determined by the decay pattern of gluinos.
If $m_{\tq} < m_{\tg}$, squarks, like
sleptons, decay\cite{CAS,BBKT}
to charginos and neutralinos. The important thing to remember is that sfermions
dominantly decay via
the two-body mode.

The various partial decay widths can be easily computed using the Lagrangian we
have described above. Numerical
results may be found in the literature for both sleptons~\cite{BBKMT} and
squarks~\cite{BBKT} and will not be repeated
here. The following features, however, are worthy of note:

\begin{itemize}

\item The electroweak decay rates are $\sim  \alpha m_{\tf}$ corresponding to
lifetimes
of about $10^{-22}(\frac{100 \  GeV}{m_{\tf}})$ seconds. Thus sfermions decay
without
leaving any tracks in the detector. We will leave it to the reader to check
that the same
is true for the decays of other sparticles discussed below.

\item Light sfermions directly decay to the LSP. For heavier sfermions, other
decays also become accessible. Decays
which proceed via the larger SU(2) gauge coupling are more rapid than those
which proceed via the smaller
U(1) coupling (Higgs couplings are negligible). Thus, for $\tf_L$, the decays
to charginos dominate unless
they are kinematically suppressed, whereas
$\tf_R$ ($f \not= t$)mainly decays into the neutralino with the largest U(1)
gaugino component.

\item Very heavy sleptons (and squarks, if the gluino mode is forbidden)
preferentially decay into the
lighter (heavier) chargino ($\tf_L$ only) and the lighter neutralinos
$\tz_{1,2}$ (the heavier neutralinos
$\tz_{3,4}$) if $|\mu|$ ($m_{\tg}$) is very large. This is because $\tw_1,
\tz_{1,2}$ ($\tw_2, \tz_{3,4}$)
are the sparticles with the largest gaugino components.

\end{itemize}

{\it Top Squark Decays:} We have seen that $t$-squarks are special in that
({\it i})~the mass eigenstates
are parameter-dependent mixtures of $\tt_L$ and $\tt_R$, ({\it ii})~$\tt_1$,
the lighter of the two states
may indeed be much lighter than all other sparticles (except, of course, for
phenomenological reasons, the LSP)
even when other squarks and gluinos are relatively heavy, and ({\it iii}) top
squarks couple to charginos
and neutralinos also via their Yukawa components. As a result the decay
patterns of $\tt_1$ can differ considerably
from those of other squarks.

The decay $\tt_1 \to t\tg$ will dominate as usual if it is kinematically
allowed. Otherwise, the decays
to charginos and neutralinos, if allowed, form the main decay modes. Since
$m_t$ is rather large, it
is quite possible that the decay $\tt_1 \to t\tz_1$ is kinematically forbidden,
and $\tt_1 \to b\tw_1$
is the only tree-level two body decay mode that is accessible, in which case it
will obviously dominate.
If the  stop is lighter than $m_{\tw_1}+m_b$, and has a mass smaller than about
125~GeV (which, we will see,
is in the range of interest for experiments at the Tevatron), the dominant
decay mode of $\tt_1$
comes from the flavour-changing $\tt_1-\tc_L$ loop level mixing induced by weak
interactions~\cite{HK} and
the decay $\tt_1 \to c\tz_1$ dominates the allowed (at least four-body) tree
level decays. If $m_{\tt_1} \sim
175-225$~GeV, the three-body decays $\tt_1 \to b W\tz_1$ may be accessible,
with the two body decays
$\tt_1 \to b\tw_1$ and $\tt_1 \to t\tz_1$ still closed. How this decay, which
could be of interest
for stop  searches at Tevatron upgrades or at future $e^+e^-$ linear colliders
or the Large Hadron Collider,
compares with the loop decay is currently under investigation.

\subsection{Gluino Decays}

Since gluinos have only strong interactions, they can only decay via
\begin{equation}
\tg \to \bar{q}\tq_{L,R},  q\bar{\tq}_{L,R},
\label{eq:gluinodk}
\end{equation}
where the squark may be real or virtual depending on squark and gluino masses.
If $m_{\tg}>m_{\tq}$, $\tq_L$
and $\tq_R$ are produced in equal numbers in gluino decays (except for phase
space corrections from the
non-degeneracy of squark masses).
In this case, since $\tq_R$ only decays
to neutralinos, neutralino decays of the gluino dominate. If, as is more
likely, $m_{\tg}<m_{\tq}$, the squark
in Eq.~(\ref{eq:gluinodk}) is virtual and decays via Eq.~(\ref{eq:sfermiondk}),
so that gluinos decay via
three body modes,
\begin{equation}
\tg \to q\bar{q} \tz_i, q\bar{q'}\tw_i.
\end{equation}
In contrast to the $m_{\tg}>m_{\tq}$ case, gluinos now predominantly
decay~\cite{CAS,BBKT} into
charginos because of the large SU(2) gauge coupling, and also into the
neutralino with the largest SU(2) gauge
component. For small values of $\mu$ ($<<\mu_2$), these may well be the heavier
chargino and the heaviest
neutralino~\cite{BBKT}; if instead $\mu$ is relatively large, as
is generally the case
in SUGRA type models, the $\tw_1$ and $\tz_2$ decays of gluinos frequently
dominate.

We should also point out that our simplistic discussion above neglects
differences between
various squark masses. As we have seen in the last section, however, third
generation squarks $\tt_1$ and
$\tb_1 \sim \tb_L$
may in fact be substantially lighter than the other squarks. It could even
be~\cite{NOJ} that
$\tg  \to \bar{b}\tb_1$ and/or $\tg \to \bar{t}\tt_1$ are the only allowed
two-body decays of the gluino
in which case gluino production will lead to  final states with very large
$b$-multiplicity, and possibly
also hard, isolated leptons from the decays of top or stop quarks. Even if
these decays are kinematically
forbidden, decays to third generation fermions may nonetheless be large because
of enhancement
of the  $\tt_1$ and $\tb_1$ propagators (recall that the decay rates roughly
depend on $\frac{1}{m_{\tq}^4}$) with
qualitatively the same effect.

Finally, we note that there are some regions of parameter space where the
radiative decay,
\begin{equation}
\tg \to g \tz_i,
\end{equation}
can be important~\cite{BTWRAD}. This decay, which occurs via third generation
squark and quark loops,
is typically enhanced relative to the tree-level decays
if the neutralino contains a large $\tilde{h}$ component (which has large
Yukawa
couplings to the top family).

\subsection{Chargino and Neutralino Decays}

Within the MSSM framework where baryon and lepton number are conserved,
charginos and neutralinos
can either decay into lighter charginos and neutralinos and gauge or Higgs
bosons, or into fermion-sfermion pairs
if these decays are kinematically allowed. We will leave it as an exercise to
the reader
to make a listing of all the allowed modes and refer the reader to the
literature~\cite{HHG,BBKMT}
for various formulae and numerical values of the branching fractions.
If these two-body decay modes are all forbidden, the
charginos and neutralinos decay via three body modes,
\begin{equation}
\tw_i \to  f\bar{f'}\tz_j, \tw_2 \to f\bar{f}\tw_1 \nonumber
\end{equation}
\begin{equation}
\tz_i \to f\bar{f} \tz_j \  or \  f\bar{f'}\tw_1,
\end{equation}
mediated by virtual gauge bosons or sfermions (amplitudes for Higgs boson
mediated decays, being proportional
to fermion masses are usually negligible).
Typically, only the lighter chargino and the neutralino $\tz_2$ decay via three
body modes, since
the decays $\tz_{3,4} \to \tz_1Z$ or $\tz_1 H_{\ell}$ and $\tw_2 \to W\tz_1$
are often kinematically accessible.
Of course if the $\tz_2$ or $\tw_1$ are heavy enough they will also decay via
two body decays: these
decays of $\tz_2$ are referred to as ``spoiler modes'' since, as we will see,
they literally spoil~\footnote{The
decay to Higgs does not yield leptons, whereas the decay to $Z$ has additional
backgrounds from SM $Z$ sources.}
the clean leptonic signal via which $\tz_2$ may be searched for.

For sfermion masses exceeding about $M_W$, $W$-mediated decays generally
dominate the three body
decays of $\tw_1$, so that the leptonic branching for its decays fraction is
essentially the same as that
of the $W$; {\it i.e. } 11\% per lepton family. An exception occurs when $\mu$
is extremely large
so that the LSP is mainly a U(1) gaugino and $\tw_1$  dominantly an SU(2)
gaugino. In this case,
the $W\tw_1\tz_1$ coupling is considerably suppressed:
then, the amplitudes for $\tw_1$ decays
mediated by virtual sfermions may no longer be negligible, even if sfermions
are relatively heavy, and the leptonic branching fractions may deviate
substantially from their canonical value of 11\%.

One may analogously expect that $\tz_2$ decays are dominated by (virtual) $Z^0$
exchange if sfermion masses
substantially
exceed $M_Z$. This is, however, not true since the $Z^0$ couples only to the
Higgsino components of the
neutralinos,
so that if either of the neutralinos in the decay $\tz_2 \to \tz_1 f\bar{f}$
has small Higgsino
components the $Z^0$ contribution may be strongly suppressed, and the
contribution from amplitudes
involving relatively heavy sfermions may be non-negligible. This phenomenon is
common in SUGRA models
where $|\mu|$ is generally much larger than the electroweak gaugino masses, and
$\tz_1$ and $\tz_2$ are,
respectively, mainly the hypercharge and SU(2) gauginos. If, in addition,
$m_{\tq} \sim m_{\tg}$, we see
from Eq.~(\ref{eq:squark}) that sleptons are much lighter than squarks, so that
the leptonic decays
$\tz_2 \to \ell \bar{\ell} \tz_1$, which lead to clean signals at hadron
colliders, may be considerably
enhanced.~\cite{BT} There are, however, other regions of parameter
space, where sleptons are relatively light, but the amplitudes from virtual
slepton exchanges
interfere destructively with the $Z^0$ mediated amplitudes, and lead to a
strong suppression
of this decay.~\cite{BCKT,MRENNA} Of course, the branching fraction for the
three-body decay is tiny
if two-body ``spoiler modes'' $\tz_2 \to Z\tz_1$ or $\tz_2 \to H_{\ell}\tz_1$
are kinematically allowed.
For basically the same reasons the decay $\tz_2 \to \tw_1 f\bar{f'}$
which is mediated by virtual $W$ exchange, even though it is
kinematically disfavoured,
can sometimes be competitive~\cite{BDDT} with the LSP decay mode of $\tz_2$.
A complete set of formulae useful for evaluating the rates for the three body
decays
of charginos and neutralinos may be found~\footnote{We warn the reader
that their notation and conventions do not match those used in these lectures,
so some care
should be exercised in transcribing these into a common notation.} in Bartl
{\it et. al.}~\cite{BFM}

Finally, we note that there are regions of parameter space where the rate for
the two body radiative decay
\begin{equation}
\tz_2 \to \tz_1 \gamma
\end{equation}
which is mediated by $f\tf$ and gauge boson-gaugino loops may be
comparable~\cite{KOM,HW} to that for the three body
decays. These are important in two different cases: ({\it i}) if one of the
neutralinos is
photino-like and the other Higgsino-like, both $Z^0$ and sfermion mediated
amplitudes are small since
the photino (Higgsino) does not couple to the $Z$-boson (sfermion),
and ({\it ii}) both neutralinos are Higgsino-like
and very close in mass (this occurs for small values of $|\mu|$); the strong
suppression of the three-body
phase space then favours the two-body decay. We mention here that neither of
these cases is particularly likely,
especially within the SUGRA framework.

\subsection{Higgs Boson Decays}

Unlike in the SM, there is no clear dividing line between the phenomenology of
sparticles and that of Higgs bosons, since as we have just seen, Higgs bosons
can also be produced via cascade
decays of heavy sparticles. Higgs boson decay patterns exhibit~\cite{HHG} a
complex dependence on model parameters.
Unfortunately, we will not have time to discuss these here, and we can only
refer the reader to the literature.
We will, therefore, confine ourselves to mentioning a few points that will be
important for later discussion.

In SUGRA models, all but the lightest Higgs scalar tend to be (but are not
always) rather heavy
and so are not significantly produced either in sparticle decay cascade decays
or directly at colliders.
Within the more general MSSM framework, the scale of their masses is fixed by
$m_{H_p}$, which is
an independent parameter.
In this limit, $H_{\ell}$ which has a mass smaller than $\sim 130$~GeV,
behaves like the SM Higgs boson, while $H_h$, $H_p$ and $H^{\pm}$ are
approximately
decoupled from vector boson pairs. The phenomenology is then relatively simple:
the decay $H_{\ell} \to b\bar{b}$
which occurs via $b$-quark Yukawa interactions dominates, unless charginos
and/or neutralinos are also
light; then, decays of $H_{\ell}$ into neutralino or chargino pairs, which
occur
via the much larger gauge
coupling, may be dominant. The invisible decay $H_{\ell} \to \tz_1\tz_1$, is
clearly the one most likely
to be accessible, and has obvious implications for Higgs phenomenology. These
supersymmetric decay modes
are even more likely for the heavier Higgs bosons, particularly if their decay
to $t\bar{t}$ pairs
is kinematically forbidden; this is especially true for $H_p$ which cannot
decay to vector boson
pairs, but also for $H_h$ since its coupling to $VV$ pairs ($V=W,Z$) is
suppressed for $m_{H_h} \agt 200$~GeV.
The decays $H_p \to H_{\ell} Z$ and $H_h \to H_{\ell}H_{\ell}$ can, of course,
be important, while
$H_h \to H_p H_p$ is usually inaccessible.
Finally, charged Higgs bosons $H^+$ mainly decay via the $t\bar{b}$ mode
unless this channel is kinematically forbidden. Then, they mainly decay via
$H^+ \to WH_{\ell}$,
or if this is also  kinematically forbidden, via $H^+ \to c\bar{s}$ or $H^+ \to
\bar{\tau}\nu$
with branching fractions depending on $\tan\beta$.

\section{Sparticle Production at Colliders}\label{sec:prod}

Since $R$-parity is assumed to be conserved, sparticles can only be pair
produced by collisions of ordinary
particles. At $e^+e^-$ colliders sparticles (such as charged sleptons and
sneutrinos,
squarks and charginos) with significant couplings to either the photon or the
$Z$-boson
can be produced via $s$-channel $\gamma$ and $Z$ processes,
with cross sections comparable with $\sigma(e^+e^- \to \mu^+\mu^-)$, except for
kinematic and statistical factors. Selectron and electron sneutrino production
may also
occur via $t$-channel neutralino and chargino exchange, while sneutrino
exchange in the $t$-channel
will contribute to chargino pair production.
Neutralino production, which proceeds via $Z$ exchange in the $s$-channel and
selectron exchange
in the $t$ and $u$ channels, may be strongly suppressed if the neutralinos are
gaugino-like
and selectrons are relatively heavy. Cross section formulae as well as
magnitudes of
the various cross sections may be found {\it e.g.\/} in Baer {\it et.
al.}~\cite{BBKMT}

\begin{figure}
\centerline{\psfig{file=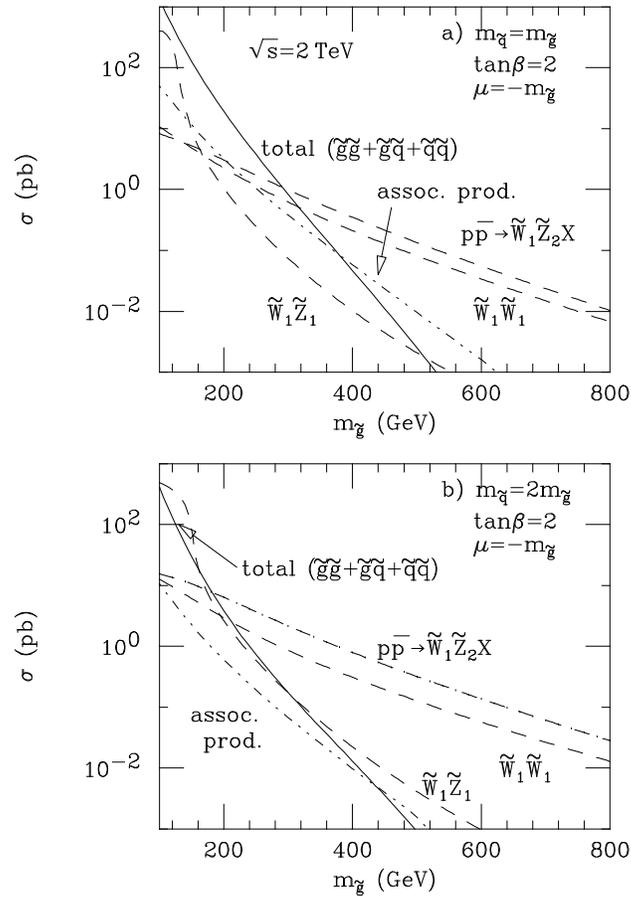,height=13cm,angle=90}}
\caption[]{ Total cross sections for various sparticle production
processes by $p\bar{p}$ collisions at $\protect\sqrt{s}=2$~TeV.}
\label{figtevcs}
\end{figure}
At hadron colliders, the situation is somewhat different. Since sparticle
production is a high $Q^2$
process, the underlying elementary SUSY process is the inelastic collision of
quarks and gluons inside the
proton.~\cite{Sterman} In other words, it is the partonic cross section that is
computable within
the SUSY framework. This cross section is then convoluted with parton
distribution functions to obtain
the inclusive cross section for SUSY particle production. Thus, unlike at
electron-positron colliders, only a fraction of the total
centre of mass energy is used for sparticle production. The balance of the
energy is contained in the
underlying low $p_T$ event which only contaminates the high $p_T$ signal of
interest.

Squarks and gluinos, the only strongly interacting sparticles, have the largest
production cross sections
unless their production is kinematically suppressed.
These cross sections~\cite{SQGLPROD} are completely determined in terms of
their
masses by QCD and do not depend on the details of the supersymmetric model. QCD
corrections to these
have also been computed.~\cite{ZERW}
Squarks or gluinos can be also
be produced~\cite{ASSPROD} in association with charginos or neutralinos via
diagrams
involving one strong and one electroweak vertex. Finally, $\tw_i$ and $\tz_j$
can be produced by $q\bar{q}$
annihilation via processes with $W$ or $Z$ exchange in the $s$-channel, or
squark exchange in the $t$ (and, for
neutralino pairs only, also the $u$) channel.

\begin{figure}
\centerline{\psfig{file=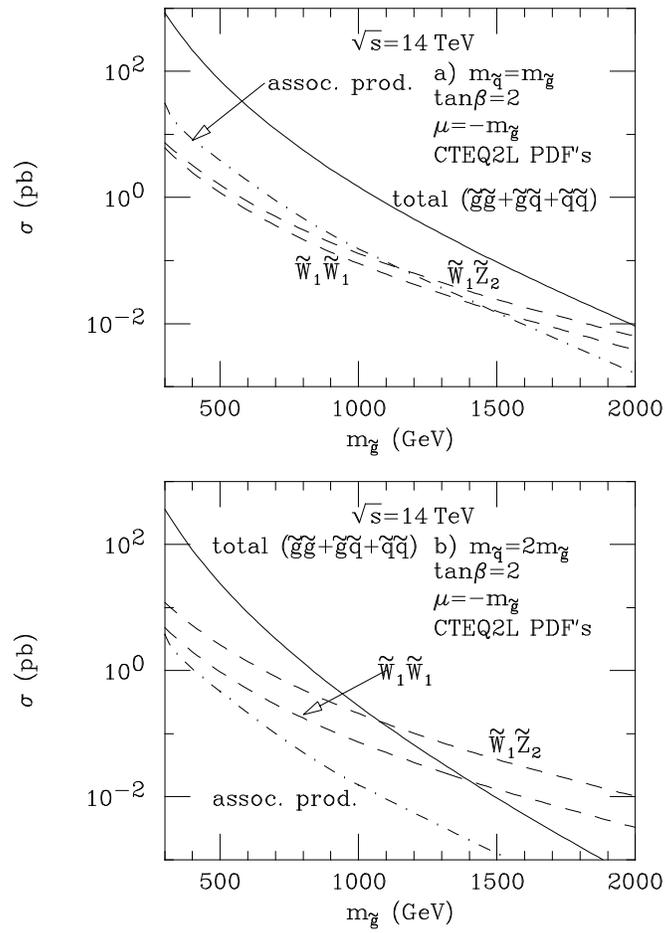,height=13cm,angle=90}}
\caption[]{Total cross sections for various sparticle production
processes by $pp$ collisions at $\protect\sqrt{s}=14$~TeV.}
\label{figlhccs}
\end{figure}

The cross sections for various processes at a 2~TeV $p\bar{p}$ collider
(corresponding to the Main Injector (MI)
upgrade of the Tevatron) are illustrated in Fig.~\ref{figtevcs},
while those for a 14~TeV $pp$ collider (the
recently approved LHC) are shown in Fig.~\ref{figlhccs}.
We have illustrated our results for
({\it a})~$m_{\tq}=m_{\tg}$, and ({\it b})~$m_{\tq}=2m_{\tg}$ and
fixed other parameters at some representative values shown. These figures help
us decide what
to search for. While squarks and gluinos are the obvious thing to focus the
initial search on, we
see from Fig.~\ref{figtevcs} that at even the MI (and certainly at the TeV33
upgrade being envisioned for
the future), the maximal reach is likely to be obtained via the electroweak
production of charginos
and neutralinos, provided of course that
their decays lead to detectable signals.~\footnote{This conclusion crucially
depends on the validity of the gaugino mass unification condition
Eq.~(\ref{eq:gaugino}).}
In contrast, we see from Fig.~\ref{figlhccs} that gluino (and, possibly,
squark) production processes offer the
best opportunity for SUSY searches at the LHC for gluino masses up to 1~TeV
(recall that
this is roughly the bound from fine-tuning considerations~\cite{FINE}) even if
squarks are very heavy.
\section{Simulation of Supersymmetry Events}\label{sec:sim}

Once produced, sparticles rapidly decay into other sparticles until the decay
cascade terminates
in a stable LSP. It is the end products of these decays that are detectable in
experiments and which,
it is hoped, can provide experimental signatures for supersymmetry. The
evaluation of these
signatures obviously entails a computation
of the branching fractions for the decays of all the sparticles, and further,
keeping
track of numerous cascade decay chains for every pair of parent sparticles.
Many groups have generated
computer programs to calculate these decay processes. For any set of MSSM
parameters (\ref{eq:mssm}) (or alternatively, for a SUGRA parameter set
(\ref{eq:sugra})),
a public access program known as
ISASUSY (ISASUGRA) which can be extracted from the Monte Carlo program
ISAJET~\cite{ISAJET} lists all sparticle and Higgs
boson masses as well as their decay modes along with the corresponding partial
widths and branching fractions.

Event generator programs provide the link between the theoretical framework of
SUSY which
provides, say, cross sections for final states with quarks and leptons, and the
long-lived
particles such as $\pi$, $K$, $\gamma$, $e$, $\mu$ {\it etc.} that are
ultimately detected
in real experiments. Many groups have combined sparticle production and decay
programs to create
parton level event generators which may be suitable for many purposes. More
sophisticated generators
include other effects such as parton showers, heavy flavour decays,
hadronization of gluons and quarks, a
model of the underlying event, {\it etc.} These improvements have significant
impact upon detailed
simulations of, for instance, the jets plus isolated multi-lepton signal from
squark and gluino
production at the LHC.

These lectures are not the place to discuss these generators in detail, so we
will content
ourselves with providing some information of what is available today.
ISAJET 7.14 is probably the most
comprehensive, but by no means complete, SUSY generator for simulation at
hadron colliders.
Mrenna~\cite{Mrennagen} has, very recently, compiled an independent event
generator for SUSY simulation
at hadron colliders. ISAJET and SUSYGEN~\cite{Kats} are two general purpose
generators available for
simulation of supersymmetry at $e^+e^-$ colliders that include all $2 \to 2$
production processes
and cascade decays of all sparticles. SUSYGEN includes initial state photon
radiation, and can be
interfaced to the LUND JETSET string hadronization program. Neither of these
generators currently
incorporates spin correlations or polarization of the
incoming beams (due to be included in the next release
of ISAJET), while final state decay matrix elements
are included only in SUSYGEN.
Specialized generators~\cite{STRASSLER,DIONGEN,MUR} that remedy
these defects are also available, but these
can only be used for the simulation of specific SUSY reactions with specific
final states.

\section{Observational Constraints on Supersymmetry}\label{sec:CONSTRAINTS}

The non-observation of any supersymmetric signal at either LEP~\cite{LEPSUSY}
or at the Tevatron~\cite{CDF,DZERO} provide
the most direct lower limits on sparticle masses. Indirect limits may also come
from
virtual effects of SUSY particles on rare processes ({\it e.g.\/} flavour
changing neutral currents or proton
decay) or from
cosmological considerations such as an over-abundance of LSP's resulting in a
universe that would be younger than
the age of stars. While these indirect limits can be important, they are
generally
sensitive to the details of the model: the non-observation of loop effects
could be a result of
accidental cancellation with some other new physics loops (so care must be
exercised in extracting
limits on sparticle masses), proton decay~\cite{PDK} is somewhat sensitive to
assumptions about GUT scale physics
while the cosmological constraints~\cite{COSM} can be simply evaded by allowing
a tiny violation of $R$-parity
conservation which would have no impact on collider searches. We should stress
that we do not mean
to belittle these
constraints which lead to important bounds in any {\it given}
framework (for instance, minimal SUGRA SU(5)), but should also recognise that
these bounds are
likely to be more model-dependent
than direct constraints from collider experiments. It is, however, only for
reasons of time
that we will mainly confine ourselves to direct limits from colliders.

The cleanest limits on sparticle  masses come from experiments at LEP. The
agreement~\cite{OLCH} of $\Gamma_Z$
with its expectation in the SM gives~\cite{WIDTH} essentially model-independent
lower limits of 30-45~GeV on the
masses of charginos, squarks, sneutrinos and charged sleptons whose couplings
to $Z^0$ were fixed
by gauge symmetry. These limits~\footnote{The same considerations also exclude
spontaneous $R$-violation
via a vev of a doublet sneutrino because the associated Goldstone boson sector
would then have
gauge couplings to $Z^0$ and make too large a contribution to $\Gamma_Z$.}
do not depend on how sparticles decay. Likewise, the measurement
of the invisible width of the $Z^0$ which gives the well-known bound on the
number of light
neutrino species, yields a lower limit on $m_{\tnu}$ only 2-5~GeV below
$\frac{M_Z}{2}$ if the sneutrino
decays invisibly via $\tnu \to \nu\tz_1$, even if only one of the sneutrinos is
light enough to
be accessible in $Z^0$ decays.~\footnote{Experiments searching for
neutrino-less double beta decay can
detect the recoil of the nucleus. If stable sneutrinos are the LSP and their
density is large
enough to form all of the galactic dark matter their flux would be high enough
to be detectable
via elastic scattering from nuclei in these experiments.  As a result,
sneutrinos with masses
between 12-20~GeV and about 1~TeV are excluded.~\cite{NONU} The Kamiokande
experiment~\cite{KAM}, from a non-observation
of high energy solar neutrinos produced by the annihilation of gravitationally
accumulated sneutrinos
in the sun exclude 3~GeV$\leq m_{\tnu} \leq $25~GeV. These limits, when
combined with the LEP
bounds clearly disfavour the sneutrino as the stable LSP.} In contrast, the
bounds on neutralino masses
are very sensitive to the model parameters because for large $|\mu|$, as we
have  already pointed out,
the neutralino may be dominantly a gaugino with strongly suppressed couplings
to the $Z^0$.

LEP experimentalists also perform direct searches for sparticles whose decays
frequently lead to extremely
characteristic final states.~\cite{CHEN} For instance, slepton (squark) pair
production
followed by the direct decay of the sfermion
to the LSP leads to a pair of hard, acollinear leptons (jets) together with
$\pslt$.
Chargino production can lead to events with
acollinear jet pairs, a lepton + jet + $\pslt$ and also acollinear leptons +
$\pslt$. Such event topologies
are very distinctive and do not occur in the SM. Thus the observation of just a
handful of such events
would suffice to signal new physics. The non-observation of SUSY signals in LEP
experiments thus
implies a lower bound on the masses of sfermions and charginos very close to
the kinematic limit.
The reactions $e^+e^- \to \tz_1\tz_2$, $\tz_2\tz_2$ can also lead to similarly
characteristic
final states. As explained above, non-observation of such signals do not lead
to bounds on neutralino
masses, but do serve to exclude regions of SUSY parameter space.~\cite{ZINO}

\begin{figure}
\centerline{\psfig{file=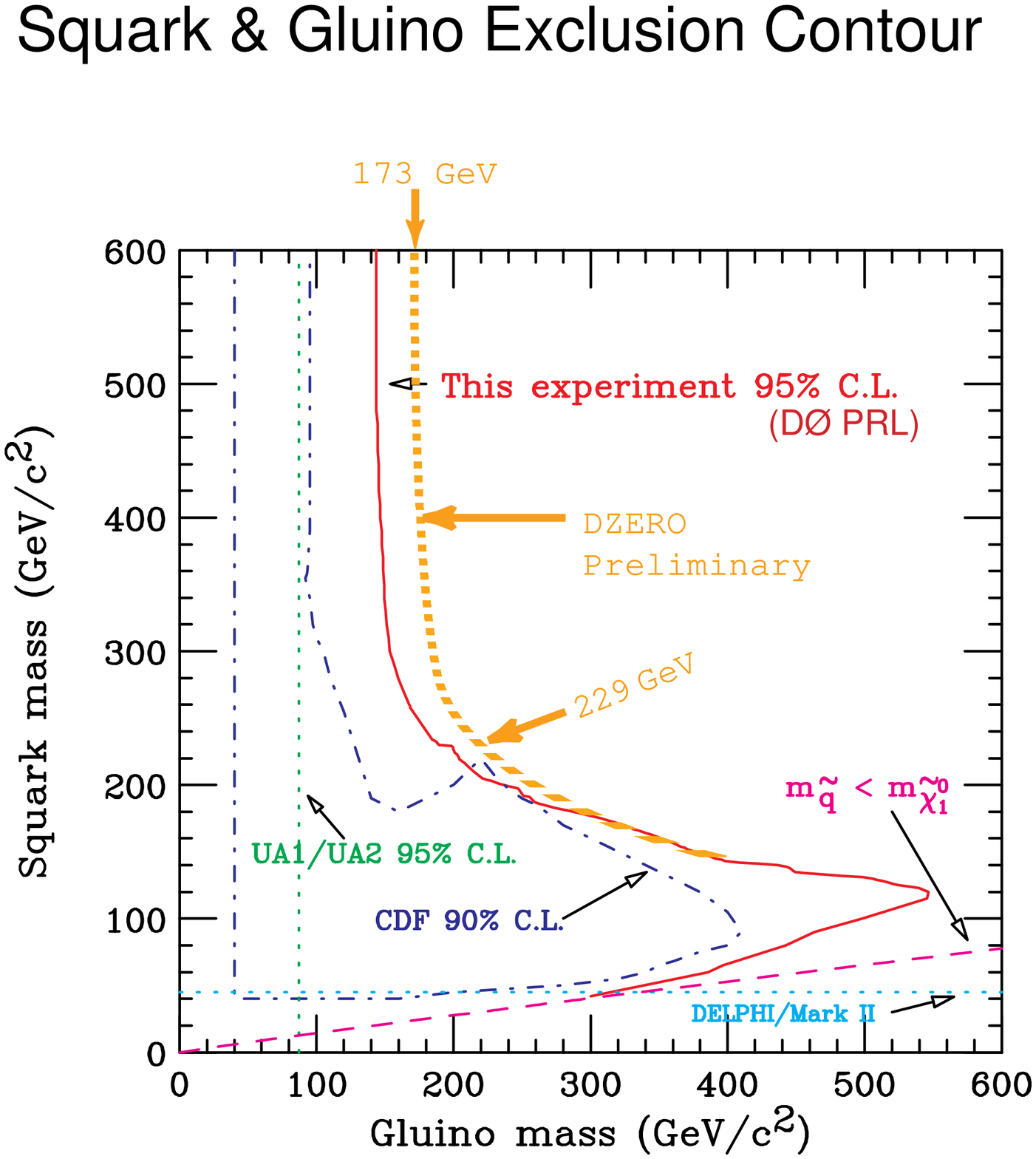,height=10.5cm}}
\caption[]{Regions of the $m_{\tg}\ vs.\ m_{\tq}$ plane excluded
by searches for $\eslt + jets$ events at various colliders, for $\tan\beta=2$
and $\mu=-250$~GeV. This figure shows the latest results from the D0
experiment and is taken from Claes.~\protect\cite{DZERO} }
\label{figd0gl}
\end{figure}
Although LEP experiments have resulted in a lower bound $\sim \frac{M_Z}{2}$ on
the squark mass,
the search for strongly interacting sparticles is best carried out at hadron
colliders by searching
for $\eslt$ events from $\tq\tq$, $\tg\tq$ and $\tg\tg$  production. The final
states from the cascade decays
of gluinos and squarks leads to events consisting of several jets plus possibly
leptons and $\eslt$.
For an integrated luminosity of about 10-20~$pb^{-1}$ on which the analyses of
the Run IA of the
CDF and D0 experiments
are based, the classic $\eslt$ channel offers the best hope for detection of
supersymmetry. The non-observation
of $\eslt$ events above SM background expectations (after cuts to increase the
signal relative to background)
has enabled the D0 collaboration~\cite{DZERO} to infer a limit of 173~GeV on
$m_{\tg}$, improving
on the published CDF limit of about 100~GeV. The region of the
$m_{\tq}-m_{\tg}$ plane excluded by these
analyses depends weakly on other SUSY parmeters and is shown in
Fig.~\ref{figd0gl} for $\mu=-250$~GeV
and $\tan\beta=2$. We see that the lower bound on the mass improves to
229~GeV if squarks and gluinos
have the same mass.
Since then, the CDF and D0 experiments have collectively accumulated about
150~$pb^{-1}$ of integrated luminosity, and should begin to be sensitive to
interesting multilepton signatures
which we will discuss when we address prospects for SUSY searches in the
future.

Before  closing this section, we briefly remark about potential constraints
from ``low  energy'' experiments, keeping
in mind that these may  be sensitive to model assumptions. As discussed in the
lectures by Hewett~\cite{joanne}
the measurements of the inclusive $b  \to s\gamma$ decay by the CLEO
experiment~\cite{CLEO} and its agreement
with SM expectations constrain the {\it sum} of SUSY contributions to this
process.~\footnote{A very recent analysis~\cite{ALI}
suggests that the theoretical error is about twice that assumed in many
analyses; this will somewhat relax the
restrictions that have been claimed in the literature.} Supersymmetry also
allows for new sources of
CP violation~\cite{CPV}
in gaugino masses or $A$-parameters. These phases, which must be smaller than
$\sim 10^{-3}$ in order that
the electric dipole moment of the neutron not exceed its experimental bound,
are set to zero in the MSSM. We will
leave it to Hewett to discuss the implications of these novel sources of CP
violation.

Finally, we note that because SUSY, unlike technicolour, is a decoupling
theory,~\cite{Paul} the agreement of the LEP
data with SM expectations is not hard to accommodate. We just have to make the
sparticles heavier than 100-200~GeV.
But by the same token, it is not easy~\cite{KANE} to accommodate the observed
deviation in the value
of $R_b = \Gamma(b\bar{b}/\Gamma(Z\to hadrons)$ (and even more so~\cite{SOLA}
for $R_c$). While the data
appear to prefer a light $\tt_1$ and a light chargino, or a light $H_p$ with
large $\tan\beta$,
it seems hard to obtain a large enough effect to explain the
``anomalies''.

\section{Searching for Supersymmetry at Future Colliders and
Supercolliders}\label{sec:Future}

\subsection{$e^+e^-$ Colliders}

LEP is scheduled to enter its second phase around the end of 1995. The
energy of LEP2 is initially expected to
be about 140~GeV, but soon should be increased to beyond the $WW$ threshold.
The signals for sparticles are
much the same as discussed in the last section. The significant difference is
that while SM backgrounds can
be easily removed below the $WW$ threshold, the separation of the SUSY signal
from $W$-pair production requires
more effort. This should not be very surprising since the $W$ is a heavy
particle and its decays can lead
to both acollinear dilepton + $\eslt$ as well as $jets+\ell +\eslt$ and $jets +
\eslt$ event topologies. Another possible complication
to be kept in mind as we search for heavier sparticles is that cascade decay
channels may begin to open up. This
should not pose too much of a problem, however, since the energy is expected to
be increased in stages. Thus,
for example, one
may expect to see chargino production before the production of sleptons which
are heavy enough to decay
to charginos sets in.

Signals for sparticle production at LEP2 have been studied in great
detail~\cite{CHEN,DION,GRIVAZ} assuming
that sparticles decay directly to the LSP. Below the $WW$ threshold, they are
readily detectable in exactly
the same way as at LEP. Above that, the production of $W^+W^-$ pairs, which has
a very large cross section
$\sim$18~$pb$ (compared to 0.2~$pb$ for smuons and $\sim 10$~$pb$ for charginos
with mass about $M_W$)
is a formidable background. The situation is not as bad as it may appear on
first sight. For $WW$ events
to fake sleptons, both $W$'s have to decay to the particular flavour of
leptons, which reduces background
by two orders of magnitude. Further rejection of background may be obtained by
noting that while slepton events
are isotropic, the leptons from  $W$ decay exhibit strong backward-forward
asymmetry. Thus by selecting
from the sample of acollinear $\mu^+\mu^-$ events those events where the fast
muon in the hemisphere
in the $e^-$ beam direction has the opposite sign to that expected from a muon
from $W$ decay, it is
possible to reduce the background by a factor of five, with just 50\% loss of
signal.

The strategy for charginos is more complicated~\cite{GRIVAZ} and will not be
detailed here. We will
only mention that here the clean environment of electron-positron colliders
plays a
crucial role. The idea is to make use of the kinematic differences between the
two-body decay
of the $W$ into a massless neutrino, and the three body decay of the chargino
into the massive LSP.
Using the cuts detailed in Ref.~\cite{GRIVAZ}, it should be possible to detect
charginos up to within
a few GeV from the kinematic limit in the mixed lepton plus jet channel.
Neutralino signals, as we
should by now anticipate, are sensitive to model parameters. A recent
analysis~\cite{BAEREE} within the framework
of the SUGRA models describes strategies to optimize these signals, and also
separate them from  other SUSY
processes.

Higher energy electron-positron colliders will almost certainly be linear
colliders, since synchroton
radiation loss in a circular machines precludes the possibility of increasing
the machine energy significantly
beyond that of LEP2. Several laboratories are evaluating the prospects
for construction of a
300-500~GeV collider, whose energy may later be increased to 1~TeV, or more:
these include the Next Linear
Collider (NLC) program in the USA, the Japanese Linear Collider (JLC) program
in Japan, the TESLA and CLIC
programs in Europe, and VLEPP in the former Soviet Union.
The search for the lightest charged sparticle, be it the chargino or the
slepton
(or perhaps the $\tt_1$) should proceed along the same
lines~\cite{JLC,MUR,CONF} as at LEP2 and discovery should be
possible essentially all the way to the kinematic limit. Of course, because
production cross sections
rapidly decrease with energy, a luminosity of 10-30~$fb^{-1}/yr$ will be
necessary.
For the more massive
sparticles, cascade decays need to be incorporated, and only relatively
preliminary work\cite{BBKMT} exists
on this. Nonetheless, all indications are that at these facilities discovery of
any massive new
particles with electroweak couplings will not pose serious difficulties. A
machine with a centre of
mass energy of about 700-1000~GeV should be able to search for charginos up to
350-500~GeV, and so,
assuming the gaugino mass unification condition, will cover the parameter space
of weak scale supersymmetry.

It is also worth
mentioning that one can exploit the availability of polarized beams to greatly
reduce
SM backgrounds: for example, the cross section for $WW$ production which is
frequently the major
background is tiny for a right-handed electron beam. While the availability of
polarized beams
and the clean environment of $e^+e^-$ collisions clearly facilitates the
extraction of the signal,
we will see later that these capabilities play a really crucial role
for the determination of sparticle properties which, in turn, serves to
discriminate
between models.

Before closing, we should mention that $e^+e^-$ colliders are ideal facilities
to search for Higgs
bosons.\cite{SOPC} At LEP2, one can typically search for Higgs bosons with a
mass up to about $\sqrt{s}-100$~GeV;
An $e^+e^-$ collider operating at 300~GeV would be virtually guaranteed to find
one of the Higgs bosons
if the MSSM framework, with its weakly coupled Higgs sector, is correct,
although it may not be possible
to distinguish this from the Higgs boson of the SM. In contrast, we will see
that at the LHC, the discovery
of the Higgs boson cannot be guaranteed even with relatively optimistic (but
not unrealistic) assumptions
about detector capabilities.

\subsection{Future Searches at Hadron Colliders}

{\it Tevatron Upgrades}

The CDF and D0 experiments have together already collected an integrated
luminosity of about
150~$pb^{-1}$ and are each expected to accumulate $\sim 100$~$pb^{-1}$ by the
end of the current run,
to be compared with 10-20~$pb^{-1}$ for the data set on which
the $\eslt$ analyses described
in the last section were based. It is thus reasonable to explore whether an
analysis
of this data can lead to other signatures for supersymmetry.
Of course,
the size of the data
sample will increase by yet another order of magnitude after about a year of MI
operations and, by
significantly more, if the TeV33 upgrade, with its design luminosity of $\sim
10~fb^{-1}/yr$, comes to pass.

{\it Gluinos and Squarks:}
While the increase in the data sample will obviously result in an increased
reach via the $\eslt$
channel~\cite{KAMON},
we have already seen that the cascade decays of gluinos and squarks lead to
novel signals
($n$ jets plus $m$ leptons plus $\eslt$) via which one might be able to
search for SUSY. Since the gluino is a Majorana particle, it decays
with equal likelihood to positive or negative charginos: the leptonic decays of
the chargino can then
lead to events with two, isolated, same-sign (SS) charged
leptons~\cite{BKP,BGH,BTW} together with jets plus $\eslt$.
\begin{figure}
\centerline{\psfig{file=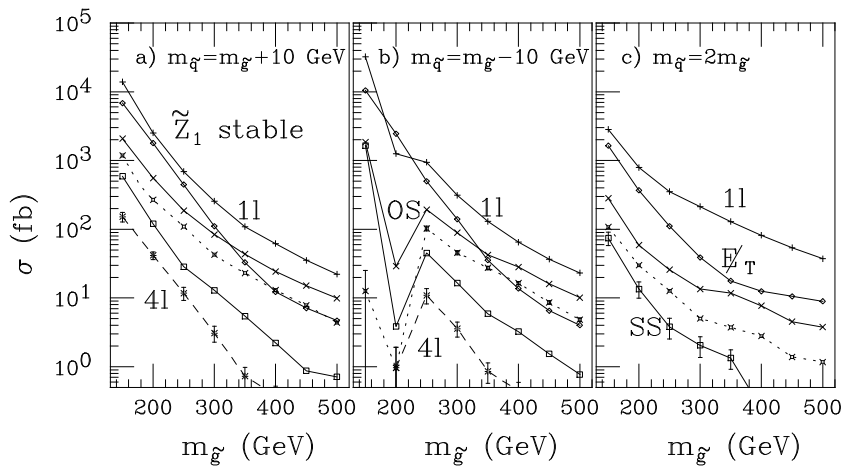,height=7cm,angle=90}}
\caption[]{Cross sections (in $fb$) at the Tevatron
($\protect\sqrt{s}=1.8$~TeV) for
various event topologies after cuts described in Ref.~\protect\cite{RPV} from
which
this figure is taken. We take $\mu=-m_{\tg}$, $\tan\beta=2$,
$A_t=A_b=-m_{\tq}$ and $m_{H_p}=500$~GeV. The $\eslt$ cross sections are
labelled with diamonds, the 1-$\ell$ cross sections with crosses, the
$\ell^+\ell^-$
cross sections with x's and the SS ones with squares.
The dotted curves are for the $3\ell$ cross sections while the dashed
curves show the cross sections for $4\ell$ events.
For clarity, error bars
are shown only on the lowest lying curve; on the other curves, these error
bars are significantly smaller. We note that the $m_{\tg}=150$~GeV case
in $b$ is already excluded by the LEP constraints on the $Z$ width, since
in this case the sneutrino mass is just 26~GeV.}
\label{figrpv}
\end{figure}
If one of the charginos is replaced by a leptonically decaying neutralino,
trilepton event topologies
result. While other topologies will also be present, the SS and $3\ell$ events
are especially
interesting because (after suitable cuts~\cite{RPV,BKTRAP}) the SM {\it
physics} backgrounds~\footnote{In
addition, there are always detector-dependent instrumental backgrounds from
misidentification
of jets or isolated pions as leptons, mismeasurement of the sign of the lepton
charge {\it etc.} that
a real experimentalist has to contend with.} are estimated to
be 2.7~$fb$ and 0.7~$fb$, respectively, for $m_t=175$~GeV.
The corresponding signal cross sections, together
with the cross sections in other channels, are illustrated in
Fig.~\ref{figrpv} which has been
obtained using ISAJET 7.13. It includes signals from all sparticle sources, not
just
gluinos and squarks. We see that while the cross sections in the clean $3\ell$
and SS event topologies are indeed tiny,
Tevatron experiments should just about be reaching the sensitivity to probe
SUSY via these channels.

{\it Charginos and Neutralinos:}
The electroweak production of charginos and neutralinos, we have seen, offers
yet another channel
for probing supersymmetry, the most promising of which is the hadron-free
trilepton signal from
the reaction $p \bar{p} \to \tw_1\tz_2 X$, where both the chargino and the
neutralino decays leptonically.
In fact, we saw in Fig.~\ref{figtevcs} that for very large integrated
luminosities, this
channel potentially  offers the maximal reach for supersymmetry (since the OS
dilepton signal from
$\tw_1 \overline{\tw_1}$ production suffers from large SM backgrounds from $WW$
production).
It was first  emphasized by Arnowitt and Nath~\cite{AN} that, with an
integrated luminosity
of $\sim 100~pb^{-1}$,  this signal would be observable at the Tevatron even if
resonance production of $\tw_1\tz_2$ is suppressed. A subsequent
analyses~\cite{BT} showed that
the signal may even be further enhanced in some regions of parameter space due
to enhancements
in the $\tz_2$ leptonic branching fractions, as discussed in
Sec.~\ref{sec:decays}. Detailed
Monte Carlo studies~\cite{BKTWINO,KAMON,BCKT,MRENNA} including effects of
experimental
cuts were performed to confirm that Tevatron experiments should indeed be able
to probe charginos
via this channel. Indeed an early analysis~\cite{CDFWINO} by the CDF
collaboration, from a non-observation of this
signal, has obtained a limit on the chargino mass that essentially coincides
with the one from LEP. While
this analysis does not yet lead to an improved bound, it nonetheless shows that
Tevatron experiments will
eventually probe regions not accessible at LEP, and perhaps, even at LEP2.

This signal, which depends on the neutralino branching fractions, is
sensitive to the model parameters, and it is not possible to simply state the
reach in terms of
the mass of the chargino. For favourable values of parameters, experiments at
the MI should be
able to probe charginos heavier than 100~GeV, corresponding to
$m_{\tg} \agt 300-350$~GeV (at TeV33, up to
$\sim$500-600~GeV
where the two-body spoiler decays of $\tz_2$ become accessible,
and for light sleptons, even up to 600-700~GeV); on the other hand,
there are other regions of parameter space where
the leptonic branching fraction of the neutralino is strongly
suppressed~\cite{BCKT,MRENNA}, and there
is no signal for charginos as light as 45-50~GeV even at the TeV33 upgrade
of the Tevatron. Thus, while this channel can
probe significant regions of the parameter space of either the MSSM or SUGRA,
the absence of any signal
in this channel will not allow one to infer a lower limit on $m_{\tw_1}$.

{\it Top Squarks:}
We have seen that $\tt_1$, the lighter of the  two top squarks, may be rather
light so that it may be pair-produced
at the Tevatron even if other squarks and gluinos are all too heavy. If its
decay to chargino is allowed, the
leptonic decay of one,  or both, stops lead to events with one, or two,
isolated leptons together with jets plus $\eslt$,
exactly the same event topologies as for the top quark search. Thus $t\bar{t}$
production is the major background~\cite{BDGGT}
to the $t$-squark search. Because $\sigma(t\bar{t}) \sim 10
\sigma(\tt_1\bar{\tt_1})$ for a top and stop
of the same mass, stop signals are detectable at the Tevatron
only if the stop is considerably lighter than the top. In the single lepton
channel it has been shown~\cite{BST,MANG} that
with an integrated luminosity
of  around 100~$pb^{-1}$, the stop signal should be detectable at the Tevatron
if $m_{\tt_1} \alt 100$~GeV, provided
$b$-jets can be adequately tagged. In the dilepton channel, the $t$-squark
signal can be separated from the top background by
searching for {\it soft} dilepton events: for instance, eliminating events with
$|p_T(\ell_1)| + |p_T(\ell_2)| + |\eslt| > 100$~GeV effectively removes the top
background if $m_t > 150$~GeV, allowing~\cite{BST}
the search for stops lighter than about 80-100~GeV without the need for
$b$-tagging.

If the chargino is heavy, the stop will instead decay via $\tt_1 \to c \tz_1$
and stop pair production will be signalled
by dijet plus $\eslt$ events, and hence looks like the squark signal, but
without any cascade decays. Again
suitable cuts will allow~\cite{BST} Tevatron experiments to probe $m_{\tt_1}
\alt 100$~GeV with 100~$pb^{-1}$ of data,
even if the LSP is relatively heavy. In fact, there is already a preliminary
analysis by the D0 collaboration~\cite{DZEROSTOP}
that excludes 60~GeV~$<m_{\tt_1}<$100~GeV if the LSP mass is smaller than
25-50~GeV.

At the MI, stop masses up to 120-130~GeV should be explorable using essentially
the same strategies.
Mrenna {\it et. al.}~\cite{MRENNA}
using cuts optimized to detect heavier stops, find that, with an integrated
luminosity of
2~$fb^{-1}$, it should be possible to explore stops as heavy as 160~GeV
if they decay via the chargino mode. They claim a reach of 200~GeV, with a data
sample of 25~$fb^{-1}$, that may be
available at TeV33. If chargino is too heavy for the decay $\tt_1 \to b\tw_1$
to be accessible but the stop mass is in the 180-250~GeV
range, we note that the three body decay $\tt_1 \to bW\tz_1$ may be
kinematically accessible. In this case, one has
to see
how this compares to the loop decay $\tt_1 \to c\tz_1$ in order to assess the
viability of this signal.

{\it Sleptons:} The best hope for slepton detection appears~\cite{SLEP} to be
via the clean OS dilepton plus $\eslt$ channel. But
even here, there is a large irreducible background from $WW$ production as well
as possible contamination of the signal
from other SUSY sources such as chargino pair production. It was concluded that
at the MI it would be very difficult
to see sleptons from off-shell $Z$ production; {\it i.e.} if
$m_{\tell} \agt 50$~GeV. Experiments at TeV33 should
probe sleptons with masses up to about 100~GeV, given an integrated luminosity
of 25~$fb^{-1}$.

{\it SUSY Searches at the LHC}

While it is certainly possible that SUSY may be discovered at an upgraded
Tevatron, there are parameter
ranges where a SUSY signal may evade detection even if sparticles are not very
heavy. There is no doubt that
in order to cover the complete parameter-space of weak scale SUSY either a
linear collider operating at
a centre of mass energy $\sim 0.5- 1$~TeV or the LHC is necessary.

We see from Fig.~\ref{figlhccs} that at the LHC, squarks and gluinos dominate
sparticle production for
gluino masses beyond 1~TeV: it is thus reasonable to focus most attention on
these although, of course,
signals should be looked for in all possible channels. For reasons of brevity,
and because the ideas
involved in LHC searches are qualitatively similar to those described above, we
will content ourselves
with just presenting an overview of the LHC reach, and refer the interested
reader to the vast amount of literature
that already exists for details.

As before, the cascade decays of gluinos and squarks result in $n$-jet plus
$m$-leptons plus $\eslt$
events~\cite{BTW2,ATLAS} where
$m=0$ corresponds to the classic $\eslt$ signal. Of the multilepton channels,
the SS and $m \geq 3$ channels
suffer the least from SM backgrounds. It is worth keeping in mind that at the
LHC, many different sparticle
chains contribute to a particular event topology, and further, that the
dominant production mechanism for
any particular channel depends on the model parameters. For instance, gluino
pair production (with each
of the gluinos decaying to a chargino of the same sign) are generally regarded
as the major source of SS dilepton events; notice, however, that the
reaction $pp \to \tb_L\bar{\tb}_L X \to t{\tw_1}^- \bar{t}{\tw_1}^+ X$
may also be a copious source of such events, since now the leptons can come
either from top or chargino decays (recall
that we had noted that $\tb_L$ may be relatively light).
It is, therefore, necessary to simultaneously
generate all possible sparticle processes in order to realistically simulate a
signal in any particular
event topology. This is possible using ISAJET. However, this raises another
issue which is especially
important at the LHC. If we see a signal in any particular channel, can we
uncover its origin? We will
return to this later, but for now, focus ourselves on the SUSY reach of the
LHC.

The ATLAS collaboration~\cite{ATLAS} at the LHC has done a detailed analysis of
the signal in the $\eslt$ as well as in
the SS dilepton channels. They found that gluinos as light as 300~GeV should be
easily detectable in the $\eslt$ channel.
Then requiring rather stiff cuts, $\eslt > 600$~GeV,
$p_T(jet_1,jet_2,jet_3)>200$~GeV, $p_T(jet_4)>100$~GeV
along with a cut $S_T>0.2$ on the transverse sphericity, they find that it
should be possible to search for
gluinos with a mass up to 1.3~TeV (2~TeV) for $m_{\tq}=2m_{\tg}$
($m_{\tq}=m_{\tg}$), assuming an integrated
luminosity of 10~$fb^{-1}$. This reach changes is altered by about $\pm300$~GeV
if the integrated luminosity
is changed by an order of magnitude. Very similar results for the reach in the
$\eslt$ channel
have also been obtained~\cite{BCPT} within the context of the
SUGRA framework, although the event selection criteria used are quite
different. In the same-sign dilepton channel,
the ATLAS collaboration~\cite{ATLAS}, concludes that the reach of the LHC will
be 900-1400~GeV (for $m_{\tq}=2m_{\tg}$)
or 1100-1800~GeV (for $m_{\tq}=m_{\tg}$), where the lower (higher) number
corresponds to a luminosity of 1~$fb^{-1}$
(100~$fb^{-1}$). Prospects for SUSY detection in the SS and other multilepton
channels (with or without real $Z$ bosons) have also
been discussed~\cite{BTW2} by other authors. An analysis of the multilepton
signals within the context of SUGRA models is in progress.
Preliminary results from this analysis~\cite{BCPT2} indicate that the SUSY
reach in the $1\ell$ channel may extend
beyond that in the $\eslt$ channel. While this may appear somewhat surprising
at first sight because this channel
is plagued by large backgrounds from $W \to \ell\nu$ and $t\bar{t}$ production,
these authors have exploited the
presence of hard jets and large $\eslt$ in SUSY events to devise cuts that
reduce this background to a manageable
level.

Within the SUGRA framework, the LHC should, in the clean  trilepton channel, be
able~\cite{LHCWINO,PHD}
to probe $\tw_1\tz_2$ production all the way up to
where spoiler modes of $\tz_2$ become accessible if $\mu < 0$ and $\tan\beta$
is not too large. Then,
it is possible to find a set of cuts that cleanly separate the $\tw_1\tz_2$
event sample from both
SM backgrounds as well as other sources of SUSY events. This will prove
important later. For positive
values of $\mu$, signals are readily observable for rather small and large
values of $m_0$; in the intermediate
range 400~GeV~$ \alt m_0 \alt 1000$~GeV, this signal is suppressed because of
the suppression of the leptonic $\tz_2$
branching fraction emphasized earlier.

As at Tevatron upgrades, the OS dilepton channel offers~\cite{AMET,SLEP} the
best opportunity for slepton searches.
At the LHC, it should be possible to detect sleptons up to about 250-300~GeV,
although excellent jet vetoing
capability will be needed to detect the signal for the highest masses.

The SUSY reach at various possible future facilities is summarized in the
Table.
\begin{table}[t]
\small
\caption[] {Estimates of the discovery
reach of various options of future hadron colliders. The signals have
mainly been computed for negative values of $\mu$. We expect that the reach
in especially the $all \to 3\ell$ channel will be sensitive to the sign of
$\mu$.}

\vskip 0.4cm
{}~\footnotesize
\begin{tabular}{c|cccc}
&Tevatron II&Main Injector&TeV33&LHC\\
Signal&0.1~fb$^{-1}$&1~fb$^{-1}$&10~fb$^{-1}$&10~fb$^{-1}$\\
&1.8~TeV&2~TeV&2~TeV&14~TeV\\
\hline
$\eslt (\tq \gg \tg)$ & $\tg(210)/\tg(185)$ &
$\tg(270)/\tg(200)$ & $\tg(340)/\tg(200)$  & $\tg(1300)$ \\

$l^\pm l^\pm (\tq \gg \tg)$ & $\tg(160)$ & $\tg(210)$ &
$\tg(270)$  & \\

$all \rightarrow 3l$ $(\tq \gg \tg)$ & $\tg(180)$ &
$\tg(260)$ & $\tg(430)$ & \\

$\eslt (\tq \sim \tg)$ & $\tg(300)/\tg(245)$ &
$\tg(350)/\tg(265)$ & $\tg(400)/\tg(265)$ &
$\tg(2000)$ \\

$l^\pm l^\pm (\tq \sim \tg)$ & $\tg(180-230)$ & $\tg(320-325)$ &
$\tg(385-405)$  & $\tg(1000)$ \\

$all \rightarrow 3l$ $(\tq \sim \tg)$ & $\tg(240-290)$ &
$\tg(425-440)$ & $\tg(550^)$  &
$\stackrel{>}{\sim}\tg(1000)$ \\


$\tilde{t}_1 \rightarrow c \tz_1$ & $\tilde{t}_1(80$--$100)$ &
$\tilde{t}_1 (120)$ & $\tt_1(150)$   &\\

$\tilde{t}_1 \rightarrow b \tw_1$ & $\tilde{t}_1(80-100)$ &
$\tilde{t}_1 (120)$ & $\tt_1(180)$   &\\

$\Theta(\tilde{t}_1 \tilde{t}_1^*)\rightarrow \gamma\gamma$ &
--- & --- & ---  & $\tilde{t}_1 (250)$\\
4~
$\tl \tl^*$ & $\tl(50)$ & $\tl(50)$ & $\tl(100)$ &
$\tl(250$--$300)$

\end{tabular}
\end{table}

Several comments  are worth noting:
\begin{itemize}

\item In some places, two sets of numbers are given for the reach. These
correspond to
results from different analyses more fully described in the review~\cite{DPF}
from where
this Table is taken. Basically, the more conservative number also requires the
signal to
be larger than 25\% of the background, in addition to exceeding the 5$\sigma$
level. Also,
the two analyses do not use the same cuts.

\item The multilepton rates in the Table are shown for negative values of $\mu$
and $\tan\beta=2$.
For other parameters, especially for $\mu > 0$, the trilepton rates may be
strongly suppressed due
to a suppression of the $\tz_2$ branching fraction discussed above. Notice
also that at TeV33,
the reach in the leptonic channels exceeds that in the $\eslt$ channel.
At the TeV33 upgrade, hadronically quiet
trilepton events may be observable all the way up to the spoiler modes for
favourable ranges
of model parameters. It is, however, important to remember that supersymmetry
may escape detection
at these facilities even if sparticles are relatively light.

\item At the LHC, gluinos and squarks are detectable to well beyond 1~TeV in
the $\eslt$ channel,
and up to 2~TeV if their masses are roughly equal. Thus the LHC should be able
to probe the complete
parameter space of weak scale SUSY, at least within the assumed framework.
Moreover, there should
be some observable signals in the leptonic channels if a signal in the $\eslt$
channel is to
be attributed to supersymmetry.

\item Tevatron upgrades will not probe sleptons significantly beyond the
reach of LEP2, whereas
the LHC reach may be comparable to that of the initial phase of linear
colliders.

\item Tevatron upgrades should be able to detect $\tt_1$ with a mass up to
120~GeV at the MI~\cite{BST}, and
up to 150-180~GeV at TeV33.~\cite{MRENNA} It has also been pointed out,
assumming that $\tt_1 \to c\tz_1$
is its dominant decay, that it should be possible~\cite{DN}to search for
$\tt_1$ at the LHC via
the two photon decay of the scalar $\tt_1\bar{\tt_1}$ bound state, in much the
same way that
Higgs bosons searches (to be discussed next) are carried out.

\end{itemize}
\setcounter{footnote}{0}
{\it Higgs Bosons:}
At the LHC, MSSM  Higgs bosons are dominantly produced~\cite{HHG} by $gg$
fusion (via loops  of quarks and squarks),
and for some parameter ranges, also via $b\bar{b}$ fusion. Vector boson fusion,
which in
the SM dominates these other mechanisms for large Higgs masses ($m\agt
600$~GeV) is generally unimportant,
since the couplings of heavy Higgs bosons to $VV$ pairs is suppressed.
Unfortunately, we do not
have much time to discuss various strategies
that have been suggested~\cite{HIGGS} for the
detection of the Higgs sector of SUSY. Over much of the parameter space, all
the Higgs
bosons except the lightest neutral scalar, $H_{\ell}$, are too heavy to be of
interest, although
for some ranges of MSSM parameters, signals from $H_h \to \gamma\gamma,
\tau\bar{\tau}, \mu\bar{\mu}, 4\ell$ and
$H_p\to \gamma\gamma, \tau\bar{\tau}, \mu\bar{\mu}$ may be observable. The two
photon decay mode is the
most promising channel for $H_{\ell}$ detection at the LHC. The region of
parameter space where
there is some signal for an MSSM Higgs boson either at the LHC or at LEP2 have
been nicely
summarized in the technical report of the CMS Collaboration~\cite{CMS},
assuming that sparticles
are too heavy to be produced via Higgs boson decays.~\footnote{This is not
necessarily a good
assumption. The branching fractions for the SUSY decays can be quite
substantial for large
regions of parameter space, and can reduce the signals via which the Higgs
bosons are usually
searched for and increase the parameter space hole referred to
below.~\cite{BISSET1} Sometimes, however, they
lead to novel signals for Higgs boson searches which can then
refill~\cite{BISSET2} some of the hole region.
Of course, for almost all cases where SUSY Higgs decays are important, it
should also be
possible to detect the sparticles at the LHC. It is only Higgs boson detection
that may be more
difficult.}
The most striking feature of their analysis is that despite optimistic detector
assumptions,
there are regions of parameter space where
{\it there may be no signal for any of the Higgs bosons either at the LHC or
at LEP2.} Part
of this hole may be excluded~\cite{joanne} by analyses of rare decays such as
$b\to s\gamma$ mentioned
earlier. It has been suggested~\cite{GUN} that Higgs boson signals may also be
detectable in this hole region via
$t\bar{t} H_{\ell}$ production
where a lepton from $t$ decay may be used to tag the event so that $H_{\ell}$
can then be
detected via its dominant $b\bar{b}$ decay. This would require efficient $b$
tagging with the high
luminosity option for the LHC. Whether this is technically possible is not
clear at this time.
It is worth remembering that Higgs boson detection would be relatively easy at
a linear collider,
the first of many examples of the complementary nature of these facilities.

\section{Beyond SUSY Discovery: More Ambitious Measurements}\label{sec:amb}

We have seen that if the minimal SUSY framework that we have adopted is a
reasonable approximation
to nature, experiments at supercolliders should certainly be able to detect
signals for physics
beyond the SM. If we are lucky, such signals might even show up at LEP2 or at
Tevatron upgrades.
We will then have to figure out the origin of these signals.
If the new physics is supersymmetry, it is likely (certain, at the LHC) that
there will simultaneously be signals in several channels. While the observation
of just
one or two of these signals would
convince the believers,
others would probably demand stronger evidence.
It is not, however, reasonable
to expect that we will immediately detect all (or even several of) the
super-partners. Thus, it is important
to think about just what information can be obtained in various experiments,
information that
will help us to elucidate the nature of the underlying physics. Towards this
end, we would like
to be able to,
\begin{itemize}

\item measure any new particle's masses and spins, and
\item measure its couplings to SM particles; these would serve to pin down its
internal quantum numbers.

\end{itemize}
More ambitiously, we may ask:
\begin{itemize}
\item Assuming that the minimal framework we have been using is correct, is it
possible to measure
the model parameters? Is it possible to actually provide tests for, say, the
minimal SUGRA framework,
and thus also test the assumptions about the physics at the GUT or Planck scale
that are an integral
part of this picture?

\item At hadron colliders, especially, where several new sparticle production
mechanisms may be simultaneously
present,~\footnote{Since the energy of the linear collider is likely to be
increased in several
steps to the TeV scale, one may hope that this will be less of a problem there.
The lighter sparticles
will be discovered  first. Knowledge about their properties thus obtained
should facilitate the
untangling of the more complex decays of heavier sparticles.} is it possible to
untangle these from one another?

\item  As mentioned in Sec.~\ref{sec:Intro}, like any other (spontaneouly
broken) symmetry,
supersymmetry, though softly broken, implies relationships between the various
couplings in the theory.
Is it possible to directly test supersymmetry by experimentally verifying these
coupling constant
relationships?

\end{itemize}

\subsection{Mass Measurements}

{\it $e^+e^-$ colliders:}
The clean environment of $e^+e^-$ colliders as well as the very precise energy
of the beam allows
for measurements of sparticle  masses. We will briefly illustrate the
underlying ideas with a simple
example. It is easy to show that the total cross section for smuon production
has the energy dependence,
\begin{displaymath}
\sigma(\tmu\bar{\tmu}) \propto (1-\frac{4{m_{\tmu}}^2}{s})^\frac{3}{2},
\end{displaymath}
and further, that the energy distribution of the daughter muon from the decay
of the smuon, assuming
only direct decays to the LSP, is flat and bounded by~\cite{ST}
\begin{displaymath}
\frac{{m_{\tmu}}^2-{m_{\tz_1}}^2}{2(E+p)} \leq E_{\mu} \leq
\frac{{m_{\tmu}}^2-{m_{\tz_1}}^2}{2(E-p)},
\end{displaymath}
with $E$($p$) being the energy (momentum) of the smuon.
We thus see that the energy dependence of the smuon cross section gives a
measure
of the smuon mass, while a measurement of the end points of the muon energy
spectrum
yields information about $m_{\tmu}$ as well as $m_{\tz_1}$.
Of course, theoretically these relations are valid for energy and momentum
measurements made
with ideal detectors without any holes and with perfect
energy and momentum resolutions. In real detectors
there would be smearing effects as well as statistical fluctutations. It has
been shown,~\cite{CHEN}
taking these effects into account, that
with an integrated luminosity of 100~$pb^{-1}$, experiments at LEP2 should be
able to determine
the smuon mass within 2-3~GeV. At the JLC, an integrated luminosity of
20~$fb^{-1}$ should enable~\cite{MUR,JLC}
the determination of the smuon and LSP masses to within 1-2~GeV. If sleptons
are heavy but charginos
light, a study of the reaction $e^+e^- \to \tw_1\overline{\tw_1} \to jj\tz_1+
\ell\nu\tz_1$
should allow the determination of $m_{\tw_1}$ and $m_{\tz_1}$ with a precision
of $\alt 3$~GeV,
both at LEP2 (where
an integrated luminosity of about 1~$fb^{-1}$ would be necessary) and at the
JLC. A good jet mass
resolution is crucial. Finally, it has also been shown~\cite{FF} that
with the availability of beam polarization at linear colliders it should be
possible
to determine squark masses with a precision of  $\sim 5$~GeV even if these
decay via
MSSM cascades: in particular, it should
be possible to determine the splittings amongst the squarks with good
precision.

{\it Hadron Colliders:}
Can one say anything about sparticle masses from SUSY signals at hadron
colliders? Despite the
rather messy environment, it is intuitively clear that this should be possible
if one can isolate
a single source of SUSY
events from both SM backgrounds as well as from other SUSY sources: a study of
the kinematics would then
yield a measure of sparticle masses within errors determined by the detector
resolution.
We have already seen that it is indeed possible to isolate a relatively
clean sample of $p\bar{p} \to  \tw_1\tz_2 \to \ell\bar{\ell}\ell' + \eslt$
events at the LHC.
The end-point of the $m_{{\ell}\bar{\ell}}$ distribution, it has been
shown,~\cite{LHCWINO,PHD} yields
an accurate measure of $m_{\tz_2}-m_{\tz_1}$. It may also be possible to
extract
relationships between the masses of charginos and neutralinos, though this is
less straightforward.

A measurement of the gluino mass would be especially important at hadron
colliders since gluinos
cannot be pair produced by tree-level processes at $e^+e^-$ colliders. The
first attempt~\cite{ASSPROD}
involved a parton-level examination of events from $\tg\tz_1$ production, with
a very hard $\eslt$ cut
to separate these events from $\tg\tg$ events. It was argued that the mass of
the hadronic system
recoiling against the missing transverse energy should yield $m_{\tg}$. The
analysis suffered from the
fact that QCD radiation as well as other SUSY processes which could contaminate
the $\tg\tz_1$ sample
were not included. It was ultimately concluded that
this reaction might be of use but only if $m_{\tg} < 350$~GeV.
Another stategy~\cite{BGH} focussed on the isolated same sign dilepton sample
from gluino pair production
(which is expected to be relatively free of SM backgrounds). Each event was
divided into two hemispheres
defined by the transverse sphericity axis, and an estimator of $m_{\tg}$
constructed. It was claimed
that $m_{\tg}$ could be measured with a precision of about 15\%. This
conclusion may be overly optimistic since
this study considered just a single SUSY source of same sign dileptons (we have
seen several sources
of such events), a single cascade decay chain ($\tg \to q\bar{q}\tw_1$, $\tw_1
\to \ell\nu\tz_1$) and
neglected effects of QCD radiation. This hemispheric separation strategy was
adopted in a recent
attempt~\cite{BCPT} to obtain $m_{\tg}$ from the $\eslt$ event sample. The
larger of the masses of the
hadronic system in the two hemispheres where events were selected requiring at
least two jets and $\eslt$
larger than a preassigned value $E_T^c$ was used as an estimator for $m_{\tg}$.
All sources of SUSY events
and QCD radiation were simulated using ISAJET. It was claimed that the gluino
mass may be measured
with a precision of 15-25\% provided $m_{\tg} \alt 800$~GeV, beyond which the
cross section becomes
too small to be able to reconstruct distributions with sufficient precision.

\subsection{Determination of Spin at $e^+e^-$ colliders}

If sparticle production occurs via the exchange of a vector boson in the
$s$-channel, it is easy
to check that the sparticle angular distribution is given by, $\sin^2\theta$
for spin zero particles,
and $E^2(1+\cos^2\theta)+m^2\sin^2\theta$ for spin half sparticles. Thus if
sparticles are produced
with sufficient boost, the angular distribution of their daughters which will
be relatively strongly
correlated to that of the parent, should be sufficient to distinguish between
the two cases.
Chen {\it et. al}~\cite{CHEN} have shown that, with an integrated luminosity of
500~$pb^{-1}$, it
should be possible to determine the smuon spin at LEP2. A similar
analysis~\cite{JLC} has been performed
for a 500~GeV linear collider.

\subsection{Testing the Minimal SUGRA GUT Framework }

We begin by recalling that all the minimal SUGRA GUT framework is determined by
just
the four parameters, $m_0$, $\mhf$, $\tan\beta$ and $A_0$ which, together with
the sign
of $\mu$ completely determine all the sparticle masses and couplings. Since the
number
of  observables can be much larger than the number of parameters, there must
exist relations
between observables which can be subjected to experimental tests. In practice,
such tests
are complicated by the fact that there are experimental errors, and further, it
may not
be possible to cleanly separate between what, in principle, should be distinct
observables; {\it e.g.}
cross sections for $\eslt$ events from $\tg\tg$, $\tg\tq$ and $\tq\tq$ sources
at the LHC.

Because of the clean experimental environment and the availability of polarized
beams, these
tests can best be done at $e^+e^-$ colliders, where we have already seen that
it is possible
to determine sparticle masses  with a precision of 1-2\%. The determination of
the selectron
and smuon masses will allow us to test their equality $m_{\te_L}=m_{\tmu_L}$,
$m_{\te_R}=m_{\tmu_R}$
at the percent level~\cite{MUR,JLC} --- the same may be done with staus, though
with a somewhat
smaller precision. This is a test of the assumed universality of slepton
masses.

A different test may be possible if both $\tell_R$ and $\tw_1$ are
kinematically accessible
and a right-handed electron beam is available. It is then possible to measure
$m_{\tz_1}$,
$m_{\tw_1}$, $\sigma_R(\tell_R\bar{\tell_R})$ and
$\sigma_R(\tw_1\overline{\tw_1})$ (note
that the chargino cross section for right-handed electron beams has no
contribution
from sneutrino exchange!). These four observables can then be fitted to the
four MSSM
parameters $\mu$, $\tan\beta$ and the electroweak gaugino masses $\mu_1$ and
$\mu_2$.
In practice~\cite{MUR,JLC}, while $\mu$ may be rather poorly determined if
the chargino is dominantly a gaugino, $\frac{\mu_1}{\mu_2}$ is
rather precisely obtained so that it should be
possible to test the gaugino mass unification condition at the few percent
level, given
50~$fb^{-1}$ of integrated luminosity.~\footnote{Feng
and Strassler~\cite{STRASSLER} have shown that, with 1~$fb^{-1}$ of data,
a test of this relation at the 20\% level may also be possible at LEP2.} It may
further be possible to determine $m_{\tnu_e}$ by measuring chargino production
with
a left-handed electron beam. This is of interest because the difference between
the squared
masses of  the sneutrino and $\tell_L$ is a direct test of the SU(2) gauge
symmetry for sleptons.
For further details and other interesting tests, we refer the reader to the
original literature.~\cite{MUR,JLC}

Analogous tests are much more difficult at hadron colliders. Nevertheless it
may be possible
to combine the data from the Tevatron and LEP experiments (again note the
complementarity
between hadron and $e^+e^-$ colliders) to test for consistency of the SUGRA
framework.
Since the signals are fixed in terms of just four parameters (and a sign) it is
convenient~\cite{BCMPT} to
display these in the $m_0-\mhf$ plane for fixed values of $\tan\beta$ and
$A_0$. That way,
various correlations become obvious. For example, it is easy to find regions
where several signals
should be simultaneously present, and at roughly predicted levels. Observation
(or non--observation)
of such signals at various facilities would serve to test these correlations
and also, perhaps, begin to
zero in on the model parameters. Of course, a determination of these is a
complex and difficult
task, and only very preliminary work has been done on this issue.

\subsection {Identifying Sparticle Production Mechanisms at the LHC}

At $e^+e^-$ colliders where the centre of mass energy is incrementally
increased, it may be reasonable
to suppose that it is unlikely (except, perhaps, for the sfermion degeneracy
expected in SUGRA type models)
that several particle thresholds will be crossed at the same time. It would
thus be possible to focus on just one new signal at a time, understand it and
then proceed to the next stage.
The situation at the LHC will, of course, be quite different. Several sparticle
production mechanisms will
be simultaneously present as soon as the machine turns on, so that even if it
is possible to distinguish
new  physics from the SM, the issue of untangling the various sparticle
production mechanisms will remain.
For example, even if we attribute a signal in the $\eslt + jets$ channel  to
sparticle production, is it
possible to tell whether the underlying mechanism is the production of just
gluinos or
of gluinos and squarks?~\footnote{Here,
we tacitly assume that squarks will not be much lighter than gluinos.}

Some progress has already been made in this direction. We have already seen
that the $\tw_1\tz_2$ source of trileptons
can clearly be isolated from other SUSY processes. The opposite sign
dilepton signal from slepton production
is probably distinguishable from the corresponding signal from
chargino pair production since the dileptons from slepton production
always have the same flavour.~\footnote{The extent to which this channel
is contaminated by other SUSY sources has not been explicitly checked.
The viability of the opposite sign, clean dilepton signal from chargino
pair production at the LHC is under investigation.}
To tell whether squarks are being produced in addition
to gluinos, at least two distinct strategies have been suggested. The first
makes use of the fact that there are more
up quarks in the proton than down quarks.
We thus expect many more $\tu_L\tu_L$ and $\tg\tu_L$ events as compared to
$\td_L\td_L$
and $\td_L\tg$ events at the LHC. As a result, any substantial production of
squarks in addition to gluinos
will be signalled~\cite{BTW3} by a charge asymmetry in the same-sign dilepton
sample: cascade decays of gluinos
and squarks from $\tg\tq$ and $\tq\tq$ events lead to a larger cross section
for positively charged same sign dileptons
than for negatively charged ones. This has since been confirmed by detailed
studies by the ATLAS collaboration~\cite{ATLAS}
where the SS dilepton charge asymmetry is studied as a function of
$\frac{m_{\tg}}{m_{\tq}}$, and shown to monotonically disappear
as this ratio becomes small.
More recently, it has been suggested~\cite{BCPT} that a study of the jet
multiplicity in the $\eslt$ sample could also
reveal the production of squarks provided one has some idea about the gluino
mass. The idea is to note that $\tq_R$,
which are produced as abundantly as $\tq_L$, frequently decay directly to the
LSP via $\tq_R \to q\tz_1$ and so lead
to only  one jet (aside from QCD radiation). In contrast, gluinos decay via
$\tg \to q\bar{q}\tw_i$ or
$\tg \to q\bar{q}\tz_i$, so that that gluino decays contain two, and frequently
more, jets from their cascade decays.
Thus the expected jet multiplicity is lower if squark production forms a
substantial fraction of the $\eslt$ sample.
Of course, since $\langle n_{jet} \rangle$ (from gluino production) depends on
its mass, some idea of $m_{\tg}$ is necessary for
this strategy to prove useful. A detailed simulation~\cite{BCPT} shows
that the mean value of the $n_{jet}$ distribution
increases by about $\frac{1}{2}$ unit, when the squark mass is increased from
$m_{\tq}=m_{\tg}$ by  about 60-80\%.

Cascade decays of gluinos and squarks could also lead to the production of
the Higgs bosons of supersymmetry. It is, therefore, interesting to ask whether
these can be detected in
the data sample which has already been enriched in SUSY events. Neutral Higgs
bosons might be detectable~\cite{BTW3}
via an enhancement of the multiplicity of central $b$-jets in the $\eslt$ or
same sign dilepton SUSY samples.
Some care must be exercised  in drawing conclusions from this because
such enhancements may also result because third generation squarks happen to be
lighter
than the other squarks.\footnote{These may be directly produced with large
cross sections or may lead to
enhancement of gluino decays to third generation fermions as discussed in
Sec.~\ref{sec:decays}.}
It has also been shown~\cite{BCPT} that it may
also be possible to reconstruct a mass bump in the $m_{b\bar{b}}$ distribution
if there is a significant branching
fraction for the decay $\tz_2 \to H_{\ell}\tz_1$ and $H_{\ell}$ is produced in
events with no other $b$-jets since
then we would have a large combinatorial background. The charged Higgs boson,
if it is light enough, may
be identifiable via the detection of $\tau$ lepton enhancements in SUSY
events~\cite{BTW3}; it should, however,
be kept in mind that such light charged Higgs bosons also contribute to the $b
\to s\gamma$ decays.

We stress that the complex cascade decay chains of gluinos and squarks may be
easier
to disentangle if we already have some knowledge about the masses and couplings
of the lighter charginos and neutralinos
that are produced in these decays. While it is indeed possible that $\tw_1$
may be discovered at LEP2 and
that its mass is determined there, it is likely that we may have to wait for
experiments
at the linear collider to be able to pin down the couplings, and, perhaps, even
for discovery of $\tw_1$
and $\tz_2$. In this case, a reanalysis of the LHC data in light of new
information that may be gained
from these experiments may prove to be very worthwhile: it may thus be
necessary to archive this data
in a form suitable for subsequent reanalaysis. Once again, we see the
complementary capabilities of $e^+e^-$
and hadron colliders.

\subsection{Direct Tests of Supersymmetry}

We have already seen that supersymmetry, like any other symmetry, implies
relationships~\footnote{These
relations are corrected by radiative corrections which are generally expected
to be smaller than a
few percent.~\cite{HIK}} between various dimensionless
couplings in the theory even if it is softly broken. For example, the
fermion-sfermion-gaugino (or, since
the Higgs multiplet also forms a chiral superfield, the Higgs-Higgsino-gaugino)
coupling is
completely determined by the corresponding gauge coupling. A verification of
the relation
would be a direct test of the underlying supersymmetry. We emphasize that such
a test would be essentially
model independent as it relies only on the underlying global supersymmetry, and
not on any details such as
assumptions about physics at the high scale or even the sparticle content. The
main complication is that
the gauginos (or the Higgs bosons and Higgsinos, or for that matter, the
sfermions) are not mass eigenstates,
so that the mixing pattern has to be disentangled before this test can be
applied. This will require an accurate
measurement of several observables which can then be used to disentangle the
mixing and also simultaneously
to measure the relevant coupling.

Feng {\it et. al.}~\cite{FMPT} have argued that such a test is best done via a
determination of chargino
properties. As we have seen, the charged gaugino and the corresponding Higgsino
can mix only if electroweak
symmetry is broken. This is the reason why the off-diagonal terms in the
chargino matrix are equal
to~\cite{Korea} $\sqrt{2}M_W\cos\beta$ and $\sqrt{2}M_W\sin\beta$,
respectively. Assuming
that the chargino is a mixture of just one Dirac gaugino and one Dirac
Higgsino, the most general mass
matrix would contain four parameters: the two diagonal elements and the two
off-diagonal ones. These
latter can always be parametrized by $\sqrt{2}M_W^{\chi}\cos\beta^{\chi}$ and
$\sqrt{2}M_W^{\chi}\sin\beta^{\chi}$.
It is the SUSY constraint on the Higgs-Higgsino-gaugino coupling that forces
$M_W^{\chi}=M_W$.
Within the MSSM framework (which we adopt for evaluating the feasibility of the
SUSY test), for parameters such that
the chargino is a substantial mixture of the gaugino and Higgsino, both
charginos should be
accessible at a 500 GeV linear collider. A determination of four quantities,
chosen to be
the masses of the two charginos along with
the total production cross section from right-handed electron beam (recall this
does not couple to the
sneutrino, so that the cross section is determined by gauge interactions) and
an appropriately
defined~\cite{FMPT} forward-backward asymmetry is thus sufficient to determine
the four entries of the chargino
mass matrix. With an integrated luminosity of 30~$fb^{-1}$, the relation
$M_W^{\chi}=M_W$ can be tested
at the 30\% level at a 500~GeV linear collider. We will refer the reader to the
original paper~\cite{FMPT} for
further details and also an analogous test (which can be done also at about the
30\% level)
for parameters such that the chargino is mainly a gaugino. Finally, if the
chargino is Higgsino-like, the lightest neutralino
is essentially degenerate with it (within the  MSSM framework).
In this case, even the observation of the chargino signal (let alone
precision tests) may be difficult because the decay products tend to be very
soft.

\section{Beyond Minimal Models}\label{sec:nonmin}

Up to now, we have confined our analysis to the MSSM framework. Even here,
we saw in Sec.~\ref{sec:mssm} that the unmanageably large number of free
parameters required us to
make additional assumptions in order to obtain tractable phenomenology. It is
clearly impractical
to seriously discuss the phenomenology of various extensions of the MSSM
framework. Here we will
merely list some of the ways in which this framework may be modified, and leave
it to the
reader to figure out the implications for phenomenology. Thinking about this
will
also  help to view our previous discussion in proper perspective.

The MSSM framework may be extended or modified in several ways.
\begin{itemize}
\item We may give up the exact universality of the gaugino masses at the GUT
scale. Threshold
corrections due to unknown GUT, and perhaps even gravitational, interactions
would certainly
yield model-dependent corrections~\cite{HALLGAUG} which preclude exact
unification.
It is also conceivable
that there is, in fact, no grand unification at all, but the observed
unification of couplings in LEP experiments
is a result~\cite{IBAN} of
string type unification; in this case, the gaugino masses need not be identical
exactly at $M_{GUT}$.
We have already  noted that even in SUGRA type models, we do not really know
the exact
scale  at which scalar  masses unify. We also remark that
the assumption of minimal kinetic energy terms is crucial
for obtaining universal scalar masses. Since the Lagrangian for supergravity
is non-renormalizable, this really is an additional~\cite{WSFN} assumption. Any
deviation from the
assumed university of scalar masses will, at the
very least, modify the
conditions  of radiative symmetry breaking.
Finally, models have also been constructed~\cite{NELS}
where SUSY breaking is a relatively low scale ($\sim 10-100$~TeV, as opposed to
$M_X$) phenomenon, so that the scalar mass
patterns will be quite different from those in minimal SUGRA.

\item $R$-parity may be explicitly broken by superpotential interactions $g_1$
and $g_2$ in Eq.~(\ref{eq:supL})
and Eq.~(\ref{eq:supB}).

\item There could be additional chiral superfields: new generations (with heavy
 neutrinos), additional
Higgs multiplets, or a right-handed sneutrino superfield. We certainly do not
need new generations
or new Higgs doublets, as they may spoil the observed unifiation of couplings.
Higgs fields in higher representations cause additional problems if they
develop a vacuum
expectation value. Higgs singlets cannot be logically excluded, and are
interesting
because they allow for new quartic Higgs boson couplings, though one would have
to understand
what keeps them from acquiring GUT or Planck scale masses. A singlet
right-handed sneutrino (note that this is {\it not} the superpartner
of the usual neutrinos) is an interesting
possibility since it occurs in SO(10) GUT models, and also, because it allows
for spontaneous breaking
of $R$-parity conservation.~\cite{VALLE}

\item Finally, we could consider models with extended low energy gauge groups
--- either left-right
symmetric models~\cite{MOHA} or models with additional $Z$ bosons.~\cite{HR}

\end{itemize}

For want of time and space, we will only qualitatively review how our earlier
discussion is altered if $R$-parity conservation
is explicitly violated.~\cite{RVIOL} In some sense, this is the minimal
extension because it does not require
the introduction of any new particles. Notice, however, that a general analysis
of this requires
the introduction of 45 new couplings: 9~$\lambda$'s, 27~$\lambda'$'s and
9~$\lambda''$'s. There are relations amongst these couplings in theories with
larger symmetries; {\it e.g.} GUTs.
As we have already discussed, many products of the baryon- and lepton-number
violating couplings
are strongly constrained. In phenomenological analyses, it is customary (and in
light of the
large number of new parameters, convenient) to assume that one of the couplings
dominates. Even so,
several of the couplings are strongly constrained. In a very nice analysis,
Barger {\it et. al.}~\cite{BARGRVIOL}
have studied the implications from various experiments --- $\beta$-decay
universality, lepton
universality, $\nu_{\mu} e$ scattering, $e^+e^-$ forward-backward asymmetries
and $\nu_{\mu}$ deep-inelastic
scattering --- for these new interactions. They find strong constraints on the
lepton-number violating
couplings, assuming~\footnote{Constraints from $\mu \to e\gamma$ or $\mu \to
3e$ decays are
much stronger if, say, both $e$ and $\mu$
number violating interactions are large.} that only one of the couplings is
non-zero: for instance, they
find that of the $\lambda$-type couplings, only $\lambda_{131}$ and
$\lambda_{133}$ can exceed 0.2  (compare
this with the electromagnetic coupling $e=0.3$) for a SUSY scale of 200~GeV,
though several $\lambda'$
and many more of the $\lambda''$ interactions can exceed this value. Dimopoulos
and Hall~\cite{DH}
have, from the upper limit on the mass of $\nu_e$, obtained a strong bound ($<
10^{-3}$) on $\lambda_{133}$.
The same argument yields significant bounds~\cite{GOD} on many of the
$\lambda'$ couplings so that of these
only $\lambda'_{112}$,$\lambda'_{121}$ and $\lambda'_{111}$ can exceed 0.2.
Generally, only the first
generation baryon number violating interactions are strongly constrained from
the non-observation of
$n-\bar{n}$ oscillations or $NN \to K\bar{K} X$.~\cite{ZWIRNER}

The reason to worry about all this is that if $R$-parity is not conserved, both
sparticle
production cross sections as well as decay patterns may be altered.
For instance, if $\lambda'$ interactions are dominant
(with $i=1$), squarks can be singly produced as resonances in $ep$ collisions
at HERA~\cite{DREINER},
or in the case of $\lambda''$ interactions, at hadron colliders.~\cite{HDE} The
production rates
will, of course, be sensitive to the unknown $R$-parity violating couplings.
Likewise, if
$R$-parity violating couplings are large compared to gauge couplings, these
$R$-violating
interactions will completely alter sparticle decay
patterns.

Even if all the $\lambda$'s
are too small (relative to gauge couplings) to significantly affect the
production and decays of sparticles (other
than th LSP),
these interactions radically alter the phenomenology because the LSP decays
visibly, so that the classic $\eslt$
signature of SUSY is no longer viable. Clearly, the phenomenology depends on
the details of the model.
Two extreme cases where the LSP decays either purely leptonically into $e$'s or
$\mu$'s and neutrinos via $\lambda$-type
couplings, or when it always decays into jets via $\lambda''$ couplings have
been examined for their impact
on Tevatron~\cite{RPV,DP} and LHC~\cite{DGR} searches for supersymmetry.
The signals, in the former case, are spectacular since
the decays of each LSP yields two leptons in addition to any other leptons from
direct decays of $\tw_1$ or $\tz_2$ produced in the  gluino or squark cascade
decays.
With an integrated luminosity of 100~$pb^{-1}$ that has already been
accumulated, experiments
at the Tevatron should be  able to probe gluinos as heavy as 500-600~GeV.
In the other case where the LSP decays purely hadronically, gluino and squark
detection is much
more difficult than in the MSSM. The reason is that the $\eslt$ signal is
greatly reduced since
neutrinos are now the only sources of $\eslt$. In fact, if squarks are heavy,
there may well be no reach in this
channel even at the Main Injector. Further, the multilepton signals from
cascade decays are also degraded because
the jets from LSP decays frequently spoil the lepton isolation. Indeed if
squarks are heavy, none of the SUSY signals
would be observable in this run of the Tevatron; even the Main Injector will
then not probe gluino masses beyond $\sim$200~GeV
(350~GeV, if $m_{\tq}=m_{\tg}$). At the LHC, attention has solely been
focussed~\cite{DGR} on the same-sign dilepton signal from gluino pair
production.
In the case where the LSP
decays purely leptonically, the gluino mass reach is in excess of 1~TeV;
in the less favourable case where the LSP decays hadronically, this signal
is again much less promising.
Sparticle detection should not be a problem in the clean environment of
$e^+e^-$ colliders, even if $R$-parity
is not conserved. In fact, LEP should be able to probe regions of parameters
not explorable
in the MSSM since signals from LSP pair production can now be
detected.~\cite{ALEPH}

\section{Concluding Remarks}\label{concl}

We have seen that experiments at the LHC should be able to explore  essentially
the whole parameter space of weak scale
supersymmetry if we require that sparticles provide the degrees of freedom
that stabilize the electroweak
symmetry breaking sector.  While experiments at Tevatron upgrades (or for that
matter even at the current
Tevatron or at LEP2) will explore substantial regions of this parameter space,
and maybe even discover
sparticles, a non-observation of any signal should not be regarded as
disheartening: the expected mass
scale is several hundred GeV up to a TeV, and so may well not be accessible
except at supercolliders. Electron-positron
linear colliders, with a centre of mass energy of 500-1000~GeV should also be
able to discover sparticles (almost
certainly so if the frequently assumed unification condition for gaugino masses
is correct). Linear
colliders are the ideal facility for the discovery and subsequent detailed
study of Higgs bosons.

The SUGRA GUT model that we have described in Sec.~\ref{sec:sugra} provides a
very attractive and economic framework.
It is consistent with known phenomenology, with grand unification and can
incorporate (though not explain) the observed
pattern of electroweak symmetry breaking. The simplest such model leads to a
degeneracy between the first
two squark generations and so is automatically consistent with constraints on
FCNC. Furthermore, because SUSY is a decoupling
theory in that virtual effects of sparticles become suppressed if their masses
are much
larger than $M_Z$, the observed agreement of the SM with LEP constraints is
simply incorporated.
These models also provide a natural candidate for cold dark matter. We have
seen, however, that these
rather predictive models are based on several assumptions about the physics at
very high scales. It is
important to keep in mind that one, or more, of these assumptions may prove to
be incorrect. This is
especially important when considering the design of future high energy physics
facilities. While it is
reasonable to use the model as a guide, it is important to examine just how
sensitively the various
signals depend on these assumptions. The important thing, however, is that
these assumptions will be testable
in future experiments. It is here that the complementary capabilities of hadron
colliders and electron-positron
linear colliders play a crucial role. The experimental verification of any of
these assumptions will provide
a window to the symmetries of physics at ultra-high energy scales.

Lest the preceeding discussion gives an impression that SUSY solves most of the
problems of particle physics,
we should remember that it addresses a single (but very important) issue: how
is electroweak symmetry broken?
It does not shed {\it any} light on the other shortcomings of the SM. For
example, SUSY has nothing to say
about the pattern of fermion masses and mixings, or the replication of
generations. While there are new sources of CP violation
in SUSY theories, it is fair to say that SUSY models do not really explain the
origin of this.
Supersymmetric theories also cause new problems not present in the SM. Why are
baryon and lepton number
conserved at low energy when it is possible to write dimension four
SU(3)$\times$SU(2)$\times$U(1) invariant
interactions
that allow for their non-conservation? Why is the supersymmetric parameter $\mu
\sim M_{Weak}$? What is the
origin of SUSY breaking and why are SUSY breaking parameters fifteen orders of
magnitude smaller than the Planck
scale? Why is the CP violation from new SUSY sources so small?

We do not know the answers to these and, probably, several other questions.
Perhaps clues to some of
these questions lie in the unknown mechanism of SUSY breaking. The measurement
of sparticle masses
(or other soft SUSY breaking parameters) in future experiments will provide
theorists with some guidance
in this regard. We should, of course, always keep open the possibility that it
is not supersymmetry, but some totally
different mechanism that is responsible for stabilizing the electroweak scale.
Only experiments can tell whether
weak scale supersymmetry is  realized in nature. What is clear, however, is
that the exploration of the
TeV scale will provide essential clues for further unravelling the nature of
electroweak symmetry breaking interactions.
We must look to see what we find.

\section*{Acknowledgements}
I thank Dave Soper for his invitation to lecture and participate
in the stimulating program of this School. I am grateful to various
students for their questions and comments. I also
thank K.~T.~Mahanthappa and his colleagues at Boulder for their generous
hospitality as well as for providing a wonderful environment
for the School. I am grateful to Howie Baer,
Vernon Barger, and especially, Manuel Drees (who also
let me know in no uncertain terms what a lousy proof-reader
I am)
for their comments on this manuscript.
I am also grateful to various colleagues for discussions
and collaborations on supersymmetry phenomenology. Without their
input, these lectures would not have been possible.
My thanks also to the members of the Phenomenology Institute,
University of Wisconsin, Madison for their wonderful hospitality during the
time that these notes were prepared. This research is supported in part
by the U.S. Department of Energy grant DE-FG-03-94ER40833.

\end{document}